\documentclass[12pt]{iopart}

\usepackage{iopams}  
\usepackage{graphicx}
\usepackage{enumerate}
\usepackage{color}

\usepackage[normalem]{ulem}  

\renewcommand\sout{\bgroup \color{red} \ULdepth=-.5ex \ULset}

\begin{document}

\title{
The Tetraquark Candidate $Z_{c}(3900)$ from Dynamical Lattice QCD 
Simulations
}

\author{Yoichi~Ikeda for HAL QCD Collaboration}

\address{
Research Center for Nuclear Physics (RCNP), Osaka University, Osaka 567-0047, Japan\\
Theoretical Research Division, Nishina Center, RIKEN, Saitama 351-0198, Japan
}
\ead{yikeda@rcnp.osaka-u.ac.jp}
\vspace{10pt}
\begin{indented}
\item[]June 2017
\end{indented}

\begin{abstract}
The structure of the tetraquark candidate $Z_{c}(3900)$, 
which was experimentally reported in $e^+ e^-$ collisions,
is studied by the s-wave meson-meson coupled-channel scattering on the lattice.
The s-wave interactions among the $\pi J/\psi$, $\rho \eta_{c}$ and $D \bar{D}^{*}$ channels are derived 
from (2+1)-flavor dynamical QCD simulations at $m_{\pi}=410$--$700$ MeV.
It is found that the interactions are dominated by 
the off-diagonal $\pi J/\psi$-$D \bar{D}^{*}$ and $\rho \eta_{c}$-$D \bar{D}^{*}$ couplings.
With the interactions obtained, the s-wave two-body amplitudes and the pole position 
in the $\pi J/\psi$-$\rho \eta_{c}$-$D \bar{D}^{*}$ coupled-channel scattering are calculated.
The results show that the $Z_{c}(3900)$ is not a conventional resonance but a threshold cusp.
A semiphenomenological analysis with the coupled-channel interaction 
to the experimentally observed decay mode is also presented to confirm the conclusion.
\end{abstract}

\maketitle

\section{Introduction}
\begin{figure}[htb]
\begin{center}
\includegraphics[width=0.70\textwidth,clip]{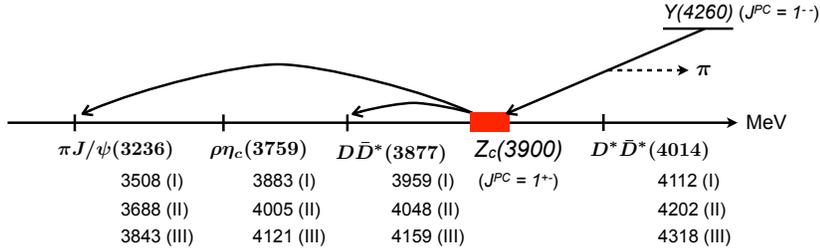}
\end{center}
\caption{
A possible decay scheme of the $Y(4260)$ through $Z_c(3900)$, 
together with the relevant two-meson thresholds of the $Z_c(3900)$ decay 
at $m_{\pi}\simeq 140$ (Expt.), $410$ (case I), $570$ (case II) and $700$ (case III) MeV.
The arrows represent the observed decay modes in the experiments~\cite{expt_BESIII,expt_Belle,expt_CLEO-c}.
}
\label{fig1}
\end{figure}

Understanding the exotic nature of hadron structures 
is one of the long-standing issues in hadron physics.
Exotic hadrons are different from the conventional well-established
quark-antiquark states (mesons) and three-quark states (baryons).
Since the discovery of the tetraquark candidate $X(3872)$~\cite{Choi:2003ue}, 
many candidates of such exotic hadrons have been experimentally reported: 
the pentaquark states $P_{c}^+(4380)$ and $P_{c}^+(4450)$ 
observed by the LHCb Collaboration~\cite{Aaij:2015tga}
and the tetraquark states $Z_{c}(3900)$ 
reported by the BESIII~\cite{expt_BESIII,Liu:2015jta}, the Belle~\cite{expt_Belle}, 
and confirmed by the CLEO-c \cite{expt_CLEO-c} Collaborations.
In particular, the resonant-like $Z_{c}(3900)$ appears as a peak 
in both the $\pi^{\pm} J/\psi$ and $\bar{D} D^{*}$ invariant mass spectra 
in the $e^{+} e^{-} \to Y(4260) \to \pi^{+} \pi^{-} J/\psi$ and $\pi D \bar{D}^{*}$ reactions:
the isospin quantum number is then identified as $I^{G}=1^{+}$, 
and the spin is favored to be $J^{PC}=1^{+-}$.
If the $Z_{c}(3900)$ is a resonance, 
it is clearly distinct from the conventional $c \bar{c}$ states:
it couples to a charmonium, yet it is charged.
So, at least four quarks, $c\bar{c}u\bar{d}$ (or its isospin partners), are involved. 
(See the level structure and the decay scheme in Fig.\ref{fig1}.)

There have been discussions 
on the internal structure of the $Z_c(3900)$ using phenomenological models.
The $Z_{c}(3900)$ has been interpreted as
a hadro-charmonium ($\pi + J/\psi$), a compact tetraquark or a $D \bar{D}^{*}$ molecule.
(e.g., Refs.~\cite{model_Voloshin, model_Cleven}) 
In these cases, the $Z_{c}(3900)$ is considered to be a resonance state.
On the other hand,
several models developed to analyse the experimental data indicate
a kinematical threshold effect (e.g., Refs.~\cite{model_Matsuki,model_Swanson}).
However, due to the lack of information of the coupled-channel interactions 
among different channels (such as $\pi J/\psi$, $\rho \eta_c$, and $D \bar{D}^{*}$), 
the predictions of those models are not well under theoretical control.
On the other hand, the direct lattice QCD studies with the standard method of temporal correlations 
show no candidate for the $Z_c(3900)$ eigenstate~\cite{LQCD_Sasa, LQCD_Lee, Chen:2014afa}, 
which indicates that the $Z_c(3900)$ may not be an ordinary resonance state~\cite{Albaladejo:2016jsg}.
Under these circumstances, it is most desirable to execute analyses
including explicitly the coupled-channel dynamics with the first-principles QCD inputs.

In this paper, we present the full account of the first lattice QCD study
to determine the nature of the $Z_c(3900)$ 
on the basis of the HAL QCD method~\cite{Ishii2007a,Aoki2010,Ishii2012,HAL2012}.   
We consider three two-body channels below $Z_{c}(3900)$
($\pi J/\psi$, $\rho \eta_c$ and $D \bar{D}^{*}$) which couple with each other.
In Ref.~\cite{Ikeda:2016zwx}, we have successfully derived 
the coupled-channel interactions faithful to the QCD S-matrix
from the equal-time Nambu-Bethe-Salpeter (NBS) wave functions on the lattice 
according to the coupled-channel formulation of the HAL QCD method~\cite{Aoki2011,Aoki2012,Sasaki2015},
and then the s-wave interaction 
has been used to calculate ideal scattering observables such as two-body invariant mass spectra
and search for the complex poles in the scattering amplitudes 
to unravel the nature of the $Z_c(3900)$.
We also present invariant mass spectra of the three-body decays 
$Y(4260) \to  \pi \pi J/\psi$ and $\pi D \bar{D}^{*}$ using the scattering amplitudes obtained in lattice QCD, 
and the results are then compared with experimental data.
It is noted here that the HAL QCD method is theoretically identical to
the L\"{u}scher's method~\cite{Luscher:1990ux} to obtain scattering amplitudes.
Numerical tests on the consistency between the both methods has been performed
in the nonresonant $\pi\pi$ channel in Refs.~\cite{Kurth:2013tua, Iritani:2015dhu} 
(See also Refs.~\cite{Iritani:2016jie} and \cite{Iritani:2017rlk}.).
We also note that the results for two-baryon systems with hyperons~\cite{Doi:2017cfx, Nemura:2017bbw, Sasaki:2017ysy, Ishii:2017xud}
obtained by the coupled-channel HAL QCD method will be compared with
future experimental data.

The paper is organized as follows.
In Sec.~\ref{HAQ_QCD_method}, 
we present the coupled-channel formula of the HAL QCD method
to calculate the scattering amplitude in the
$\pi J/\psi$-$\rho \eta_c$-$D \bar{D}^{*}$ coupled system.
Then, the numerical setup of lattice QCD simulations are summarized
in Sec.~\ref{LQCD_setup}.
The results on the coupled-channel potentials, scattering amplitude and pole position
are presented in Sec.~\ref{results_2-body}.
In Sec.~\ref{results_3-body}, we discuss the three-body decay of the $Y(4260)$
and compare our results with the experimental data.
Sec.~\ref{summary} is devoted to summary.

\section{The coupled-channel HAL QCD method}
\label{HAQ_QCD_method}
In this section, we briefly review the coupled-channel HAL QCD method~\cite{Aoki2011,Aoki2012,Sasaki2015,Ikeda:2016zwx}
by which we extract coupled-channel energy-independent potentials in hadron-hadron scatterings.
We begin with a two-hadron correlation function,
\begin{eqnarray}
C^{\alpha \beta}(\vec{r},t-t_{0}) \equiv 
\frac{1}{ \sqrt{Z_{1}^{\alpha} Z_{2}^{\alpha}} }
\sum_{\vec{x}} \left\langle 0 \right| \phi^{\alpha}_{1}(\vec{x} + \vec{r}, t) \phi^{\alpha}_{2}(\vec{x}, t) 
\overline{\mathcal{J}}^{\beta}(t=t_{0}) \left| 0 \right\rangle  ,
\label{eq:Cab}
\end{eqnarray}
where each channel is specified by $\alpha=(\pi J/\psi, \rho \eta_c, D \bar{D}^{*})$,
and $\phi^{\alpha}_{i}(\vec{y},t)$ is a local Heisenberg operator
at Euclidian time $t > t_{0}$ and the spatial point $\vec{y}$
for the meson $i \ (=1,2)$ with mass $m_{i}^{\alpha}$ in channel $\alpha$.
$\overline{\mathcal{J}}^{\beta}(t_{0})$ denotes a two-meson operator in channel $\beta$ 
located at time $t=t_{0}$.
The correlation function is normalized by the wave function renormalization factor given by $Z_{i}^{\alpha}$.
The NBS wave function $\psi^{\alpha}(\vec{r}; W_{n})$ for each scattering state specified by $n$-th energy level on the lattice is defined by
\begin{eqnarray}
\psi^{\alpha}(\vec{r}; W_{n}) =
\sum_{\vec{x}} \left\langle 0 \right| \phi^{\alpha}_{1}(\vec{x} + \vec{r}, t) \phi^{\alpha}_{2}(\vec{x}, t) \left| n \right\rangle  ,
\end{eqnarray}
and is related to Eq.~(\ref{eq:Cab}) as 
\begin{eqnarray}
C^{\alpha \beta}(\vec{r},t-t_{0}) = 
\sum_n  \psi^{\alpha}(\vec{r}; W_{n}) A_{n}^{\beta} e^{-W_n \left(t-t_0\right)}
\end{eqnarray}
with the energy eigenvalue $W_n$ of the $n$-th QCD eigenstate.
The coefficient
$A_{n}^{\beta} \equiv \langle W_n | \overline{\mathcal{J}}^{\beta}(t_{0}) | 0 \rangle$ 
is an overlap between the eigenstate and QCD vacuum 
by the insertion of the source operator $\overline{\mathcal{J}}^{\beta}(t_{0})$.

Outside a hadron-hadron interaction with the range $R$, 
the NBS wave function in each channel $\alpha$ satisfies the free Schr\"{o}dinger equation as
\begin{eqnarray}
\left(
\frac{\nabla^2}{2 \mu^{\alpha}} + \frac{\left( \vec{p}_n^{\alpha} \right)^2}{2 \mu^{\alpha}}
\right)
\psi^{\alpha}(\vec{r}; W_{n}) = 0 
~~~ \left( \left| \vec{r} \right| > R \right) ,
\end{eqnarray}
with $\mu^{\alpha}=m^{\alpha}_{1} m^{\alpha}_{2} /(m^{\alpha}_{1} + m^{\alpha}_{2})$
being the reduced mass in channel $\alpha$.
The corresponding asymptotic momentum $\vec{p}_n^{\alpha}$ defined in the center-of-mass (c.m.) frame
is related to the energy,
\begin{eqnarray}
W_{n} = 
\sqrt{ m_{1}^{\alpha} + \left( \vec{p}_{n}^{\alpha} \right)^2 }
+
\sqrt{ m_{2}^{\alpha} + \left( \vec{p}_{n}^{\alpha} \right)^2 } .
\end{eqnarray}
On the other hand, inside the interaction,
the half-off-shell T-matrix is obtained by
\begin{eqnarray}
K^{\alpha}(\vec{r}; W_{n}) \equiv
\left(
\frac{\nabla^2}{2 \mu^{\alpha}} + \frac{\left( \vec{p}_n^{\alpha} \right)^2}{2 \mu^{\alpha}}
\right)
\psi^{\alpha}(\vec{r}; W_{n})
~~~ \left( \left| \vec{r} \right| \le R \right) ,
\end{eqnarray}
from which we define the energy-independent coupled-channel potential matrix,
\begin{eqnarray}
K^{\alpha}(\vec{r}; W_{n}) 
=
\left( E^{\alpha}_{n} - H^{\alpha}_{0} \right) \psi^{\alpha}(\vec{r}; W_{n})
\equiv
\sum_{\beta} \int d\vec{r'}
U^{\alpha \beta}( \vec{r}, \vec{r'} )
\psi^{\beta}(\vec{r'}; W_{n}) .
\label{eq:cc_pot_U}
\end{eqnarray}
with $E^{\alpha}_n = (\vec{p}^{\alpha}_n)^2/2\mu^{\alpha}$ and $H^{\alpha}_{0}=-\nabla^{2}/2\mu^{\alpha}$ in channel $\alpha$.
The energy-independent coupled-channel potential $U^{\alpha \beta}(\vec{r},\vec{r'})$ guarantees that 
the S-matrix is unitary and gives the correct scattering amplitude 
until the new inelastic threshold opens~\cite{Aoki2011,Aoki2012}. 
The non-locality of the coupled-channel potential $U^{\alpha \beta}(\vec{r}, \vec{r'})$ is handled
in terms of the velocity expansion,
$U^{\alpha \beta}(\vec{r},\vec{r'}) = 
( V^{\alpha \beta}_{\mathrm{LO}}(\vec{r}) + V^{\alpha \beta}_{\mathrm{NLO}}(\vec{r}) + \cdots ) \delta(\vec{r} - \vec{r'})$, 
and we extract the leading order potential $V^{\alpha \beta}_{\mathrm{LO}}(\vec{r})$ in this study.

For example, $2 \times 2$ coupled-channel problems
(e.g., $A \leftrightarrow B$ coupled-channel with the thresholds $m_1^{A}+m_2^{A} < m_1^{B}+m_2^{B}$),
the above equation Eq.~(\ref{eq:cc_pot_U}) is solved as
\begin{eqnarray}
\fl
\left(
\begin{array}{cc}
V_{\mathrm{LO}}^{A, A}( \vec{r} ) &
V_{\mathrm{LO}}^{A, B}( \vec{r} )
\\
V_{\mathrm{LO}}^{B, A}( \vec{r} ) &  
V_{\mathrm{LO}}^{B, B}( \vec{r} )
\end{array}
\right)
\nonumber \\
\fl
~~~~~
=   
\left(
\begin{array}{cc}
K^{A}(\vec{r}; W_{1}) &
K^{A}(\vec{r}; W_{2})
\\
K^{B}(\vec{r}; W_{1}) &
K^{B}(\vec{r}; W_{2}) 
\end{array}
\right)
\cdot
\left(
\begin{array}{cc}
\psi^{A}(\vec{r}; W_{1}) &
\psi^{A}(\vec{r}; W_{2})
\\
\psi^{B}(\vec{r}; W_{1}) &
\psi^{B}(\vec{r}; W_{2}) 
\end{array}
\right)^{-1} .
\end{eqnarray}
This implies that at least the two levels of the NBS wave functions,
$\{ \psi^{A}(\vec{r}; W_{n}), \psi^{B}(\vec{r}; W_{n}) \}_{n=1,2}$
which are connected to each scattering state with a given energy $W$
in the infinite volume if $W \ge m_1^{B}+m_2^{B}$,
have to be extracted from lattice QCD simulations.
In our full $3 \times 3$ the $\pi J/\psi$-$\rho \eta_c$-$D \bar{D}^{*}$ coupled system,
$\{ \psi^{\pi J/\psi}(\vec{r}; W_{n}), ~ \psi^{\rho \eta_c}(\vec{r}; W_{n}), ~ \psi^{D \bar{D}^{*}}(\vec{r}; W_{n}) \}_{n=1,2,3}$
are needed.
Optimizing the source operator, 
$\overline{\cal J}^{\alpha}$ $(\alpha = \pi J/\psi,~ \rho \eta_c, ~ D \bar{D}^{*})$,
in  Eq.~(\ref{eq:Cab}),
one can obtain these sets of the NBS wave functions.
However, as the spatial extent $L$ gets to be large, 
it becomes difficult to identify the energy eigen-states in the simulations.
Therefore, we employ what is called the time-dependent coupled-channel HAL QCD method
to calculate the coupled-channel potential, with which the energy eigen-states need not be identified.

We introduce the hadron 4-pt correlation function $R^{\alpha \beta}$ defined by
\begin{eqnarray}
R^{\alpha \beta}(\vec{r}, t-t_{0})
\equiv
\frac{C^{\alpha \beta}(\vec{r}, t-t_{0}) }{e^{-(m_1^\alpha+m_2^\alpha)(t-t_0)}}
=
\sum_n  \psi^{\alpha}(\vec{r}; W_{n}) e^{-\Delta W^{\alpha}_n \left(t-t_0\right)}
A_{n}^{\beta} ,
\end{eqnarray}
with $\Delta W^{\alpha}_n = W_n - (m^{\alpha}_1 + m^{\alpha}_2)$.
In the nonrelativistic approximation $\Delta W^{\alpha}_n \simeq E^{\alpha}_n$,
the kinetic energy in Eq.~(\ref{eq:cc_pot_U}) is replaced with the time derivative,
so that 
$R^{\alpha \beta}$ satisfies 
the time-dependent Schr\"{o}dinger-type equation~\cite{Ishii2012,Aoki2012},
\begin{eqnarray}
\left( -\frac{\partial}{\partial t} - H^{\alpha}_{0} \right)
R^{\alpha \beta}(\vec{r}, t-t_{0}) 
=  
\sum_{\gamma} \Delta^{\alpha \gamma}
V_{\mathrm{LO}}^{\alpha \gamma}(\vec{r}) 
R^{\gamma \beta}(\vec{r}, t-t_{0})  ,
\label{t-dep_local}
\end{eqnarray}
with $\Delta^{\alpha \gamma}=e^{(m_{1}^{\alpha}+m_{2}^{\alpha}) (t-t_0)}/e^{(m_{1}^{\gamma}+m_{2}^{\gamma}) (t-t_0)}$.
The nonrelativistic approximation can be removed,
if the higher order time derivative terms associated with relativistic corrections are included, 
${\mathcal O}\bigl( (\partial_{t}^{2} / m^{\alpha}_{1,2}) (\partial_{t} / m^{\alpha}_{1,2})^{n} \bigr)$ with $n \ge 0$,
whose contributions, however, turn out to be numerically negligible 
in the present lattice setup with relatively large pion masses.
We consider $t$ sufficiently large that the inelastic states (The lowest one is $D^{*} \bar{D}^{*}$ in the current lattice QCD setup.) become negligible in $V^{\alpha \beta}$, otherwise the inelastic channels should be taken into account explicitly in the coupled-channel scattering.

To extract the s-wave coupled-channel potential 
$V^{\alpha \beta}(\vec{r}) = V_{\mathrm{LO} ~ (\ell = 0)}^{\alpha \beta}(\vec{r})$, 
we project the normalized correlation function to 
the $A^{+}_{1}$ representation of the cubic group on the lattice,
\begin{eqnarray}
R^{\alpha \beta}(\vec{r}, t-t_{0}; A_1^{+}) 
\equiv  
\frac{1}{24}
\sum_{g \in \mathcal{O}} \chi^{(A_1^{+})}(g)
R^{\alpha \beta}(g^{-1} \vec{r}, t-t_{0})  ,
\end{eqnarray}
where $g \in \mathcal{O}$ are elements of the cubic group, and
the associated characters of the $A_1^{+}$ representation are given 
by $\chi^{(A_1^{+})}(g) ~ (= 1)$.
Therefore, the s-wave coupled-channel is given by
\begin{eqnarray}
\fl
\Delta^{\alpha \beta} V^{\alpha \beta}( \vec{r} )
=
\sum_{\gamma}
\biggl[ \left( -\frac{\partial}{\partial t} - H_{0} \right)
R(\vec{r}, t-t_{0}; A_1^{+}) \biggr]_{\alpha \gamma}
\cdot
\biggl[ R(\vec{r}, t-t_{0}; A_1^{+}) \biggr]^{-1}_{\gamma \beta} .
\label{eq:s-wave_V}
\end{eqnarray}
The systematic errors originating from higher derivative terms and
the contribution from inelastic states are 
estimated by the $t$ dependence of scattering observables~\cite{Ishii2012}.

It is noted that 
some phenomenological parametrization to the on-shell K-matrix is necessary
to approximate the energy dependence of the S-matrix given by lattice QCD simulations
in the coupled-channel L\"{u}scher's method proposed 
in Refs.~\cite{Doring:2011vk, Wu:2014vma, Dudek:2014qha},
while, in the HAL QCD method, the velocity expansion is employed 
to approximate the nonlocality of the coupled-channel potentials faithful to the S-matrix of QCD.

\section{Numerical setup of LQCD simulations}
\label{LQCD_setup}
We employ (2+1)-flavor QCD gauge configurations generated by the PACS-CS Collaboration~\cite{PACS-CS2009,PACS-CS2010} on a $32^3 \times 64$ lattice with the renormalization group improved gauge action at $\beta_{\rm lat} = 1.90$ and the nonperturbatively $O(a)$-improved Wilson quark action at $C_{\rm SW}=1.715$.
The parameters correspond to the lattice spacing $a = 0.0907~ (13)$ fm and the spatial lattice volume $L^3 \simeq (2.9 ~ {\rm fm})^3$.
The hopping parameters are taken to be
$\kappa_{ud} = 0.13700, ~ 0.13727, ~ 0.13754$ for $u$ and $d$ quarks and $\kappa_s = 0.13640$ for the $s$ quark.
For the charm quark action, we employ the relativistic heavy quark (RHQ) action~\cite{RHQ_Aoki},
which is designed to remove the leading order and next-to-leading order cutoff errors associated with the heavy charm quark mass, ${\mathcal O}((m_c a)^{n})$ and ${\mathcal O}((m_c a)^{n} (a \Lambda_{\rm QCD}))$, respectively.
The RHQ action is given by
\begin{eqnarray}
S_{\mathrm{RHQ}} &=& \sum_{x,y} \bar{c}(x) D(x,y) c(y)  , \\
D(x,y) &=& \delta_{x,y} 
- \kappa_{Q} \sum_{i=1}^{3} \left[ 
( r_s - \nu \gamma_i ) U_{x,i} \delta_{x+\hat{i},y} +
( r_s + \nu \gamma_i ) U^{\dagger}_{x,i} \delta_{x,y+\hat{i}} \right] \nonumber \\
& & 
- \kappa_{Q} \left[ 
( r_t - \nu \gamma_4 ) U_{x,4} \delta_{x+\hat{4},y} +
( r_t + \nu \gamma_4 ) U^{\dagger}_{x,4} \delta_{x,y+\hat{4}} \right] \nonumber \\
& & 
- \kappa_{Q} \left[ 
c_{B} \sum_{i,j} F_{ij} \sigma_{ij} + c_{E} \sum_{i} F_{i4} \sigma_{i4} 
\right] \delta_{x,y}  .
\end{eqnarray}

In our simulation, the parameters of the RHQ action are $\kappa_Q$, $r_s$, $r_t$, $\nu$, $c_{B}$ and $c_{E}$, and the redundant parameter $r_t$ is chosen to be $ 1$.
We take the same parameters as in Ref.~\cite{Namekawa2011,Ikeda2013},
where the relativistic dispersion relation of the 1S charmonium is reproduced.
The RHQ parameters are summarized in Table~\ref{tab:RHQ_param}.

\begin{table}[htbp]
   \centering
   \begin{tabular}{ccccc}
      \hline
      \hline
$\kappa_{Q}$ & $r_s$ & $\nu$ & $c_{B}$ & $c_{E}$ \\
      \hline 
0.10959947  & 1.1881607 & 1.1450511 & 1.9849139 & 1.7819512 \\
      \hline 
      \hline
   \end{tabular}
   \caption{ Parameters of the RHQ action in our simulations.
   }
   \label{tab:RHQ_param}
\end{table}

The periodic boundary conditions are imposed on the three spacial directions, 
and the Dirichlet boundary conditions are taken for the temporal direction at $(t-t_0)/a=\pm 32$
to avoid contaminations from the opposite propagation of mesons in time.
For the sink operators, we choose the local interpolating operators,
$\phi(x) = \bar{q}(x) \Gamma q(x) ~ \left( q=u, ~ d, ~ c \right)$, where $\Gamma$ denotes a $4\times 4$ matrix acting on spinor indices.
We take $\Gamma = \gamma_5$ for  pseudo-scalar mesons ($\pi$, $\eta_c$ and $D$)
and $\Gamma = \gamma_i$ for vector mesons ($\rho$, $J/\psi$ and $\bar{D}^{*}$).
For the source operators to create the $I=1$ two-meson states, 
$\pi J/\psi$, $\rho \eta_c$ and $D \bar{D}^{*}$,
we take into account the following 
zero-momentum quark wall sources:
\begin{eqnarray}
\overline{{\cal J}}_{\pi J/\psi}(t_{0}) &=& 
\sum_{\vec x_1, \vec x_2, \vec x_3, \vec x_4}
\left[
\bar{u}(\vec x_1, t_{0}) \gamma_5 d(\vec x_2, t_{0})
\bar{c}(\vec x_3, t_{0}) \gamma_i c(\vec x_4, t_{0})
\right] , \\
\overline{{\cal J}}_{\rho \eta_c}(t_{0}) &=& 
\sum_{\vec x_1, \vec x_2, \vec x_3, \vec x_4}
\left[
\bar{c}(\vec x_1, t_{0}) \gamma_5 c(\vec x_2, t_{0})
\bar{u}(\vec x_3, t_{0}) \gamma_i d(\vec x_4, t_{0})
\right] , \\
\overline{{\cal J}}_{D \bar{D}^{*}}(t_{0}) &=& 
\sum_{\vec x_1, \vec x_2, \vec x_3, \vec x_4}
\left[
\bar{c}(\vec x_1, t_{0}) \gamma_5 d(\vec x_2, t_{0})
\bar{u}(\vec x_3, t_{0}) \gamma_i c(\vec x_4, t_{0})
\right] .
\end{eqnarray}
For the improvement of statistics, 
we repeat the measurements in Eq.~(\ref{eq:Cab}) for each gauge configuration with respect to the source position $t_{0}/a = 0$ and $32$,
and both forward and backward propagations are averaged to enhance the signal.
Throughout this study, statistical errors are evaluated by the jackknife method.

The calculated meson masses together with their physical masses 
and the number of configurations $N_{\mathrm{cfg}}$ used in the simulation
are summarized in Table~\ref{tab1}.
The two-meson thresholds relevant to this study are shown in Fig.~\ref{fig1}: 
the $\pi \psi'(3826)$ threshold is above the $D \bar{D}^{*}$ threshold
because of the heavy pion mass in our simulation.
The $\rho \to \pi \pi$ p-wave decay is not allowed 
with the current spatial extent $L \simeq 3$fm, 
so the asymptotic $\rho \eta_c$ is regarded as a well-defined two-body channel.
Pair annihilations of charm quarks are not taken into account in the simulations.
\begin{table}[tbhp]
   \centering
   \begin{tabular}{c|ccccccc}
      \hline
      \hline
         & $m_{\pi}$ & $m_{\rho}$ & $m_{\eta_c}$ & $m_{J/\psi}$ & $m_{D}$ & $m_{\bar{D}^{*}}$ & $N_{\mathrm{cfg}}$   \\
      \hline 
Expt.    & 140   &  775    & 2984    & 3097    & 1870    & 2007    & \\
      \hline 
Case I   & 411(1) &  896(8) & 2988(1) & 3097(1) & 1903(1) & 2056(3) & 450 \\
      \hline 
Case II  & 570(1) & 1000(5) & 3005(1) & 3118(1) & 1947(1) & 2101(2) & 400 \\
      \hline 
Case III & 701(1) & 1097(4) & 3024(1) & 3143(1) & 2000(1) & 2159(2) & 399 \\
      \hline 
      \hline 
   \end{tabular}
   \caption{ Meson masses in MeV units and the number of configurations used 
   in our simulations. }
   \label{tab1}
\end{table}

\section{Coupled-channel potential and structure of $Z_c(3900)$}
\label{results_2-body}
\begin{figure*}[!htb]
\includegraphics[width=0.32\textwidth,clip]{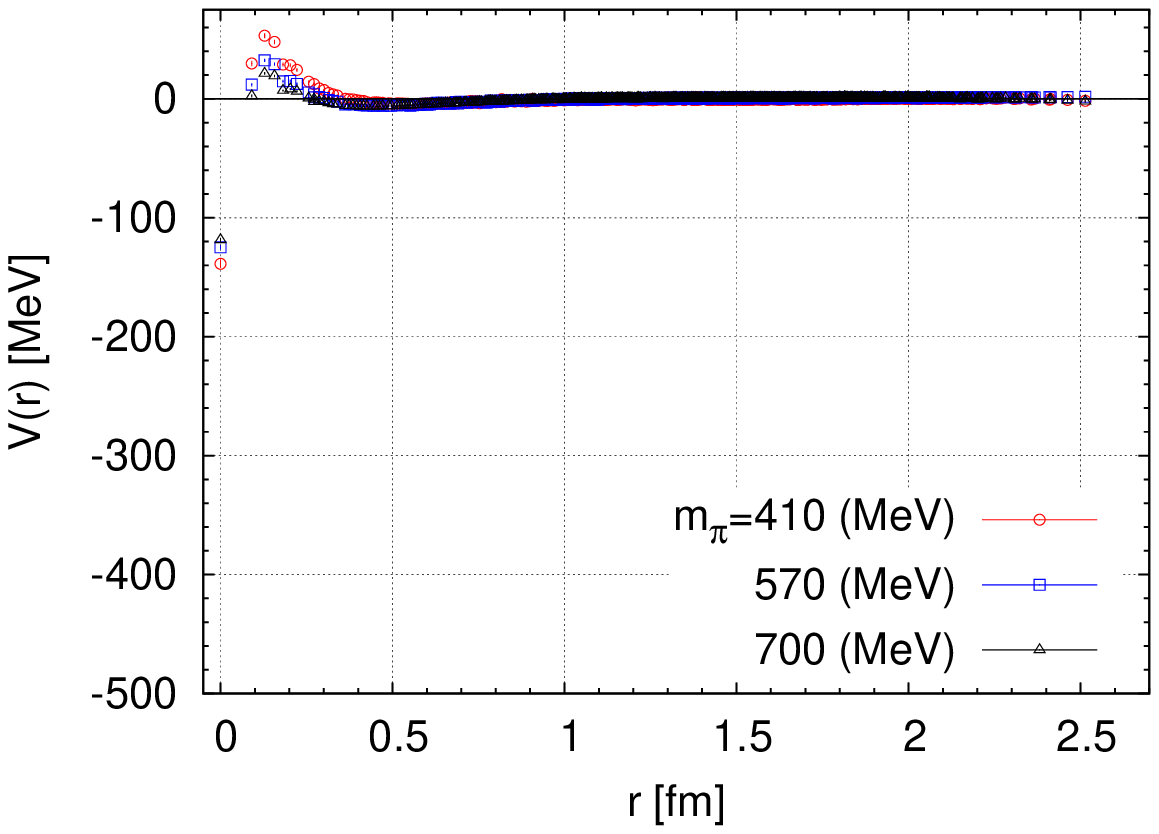}
\put(-80,50){\small (a) $V^{\pi J/\psi, \pi J/\psi}$}
\includegraphics[width=0.32\textwidth,clip]{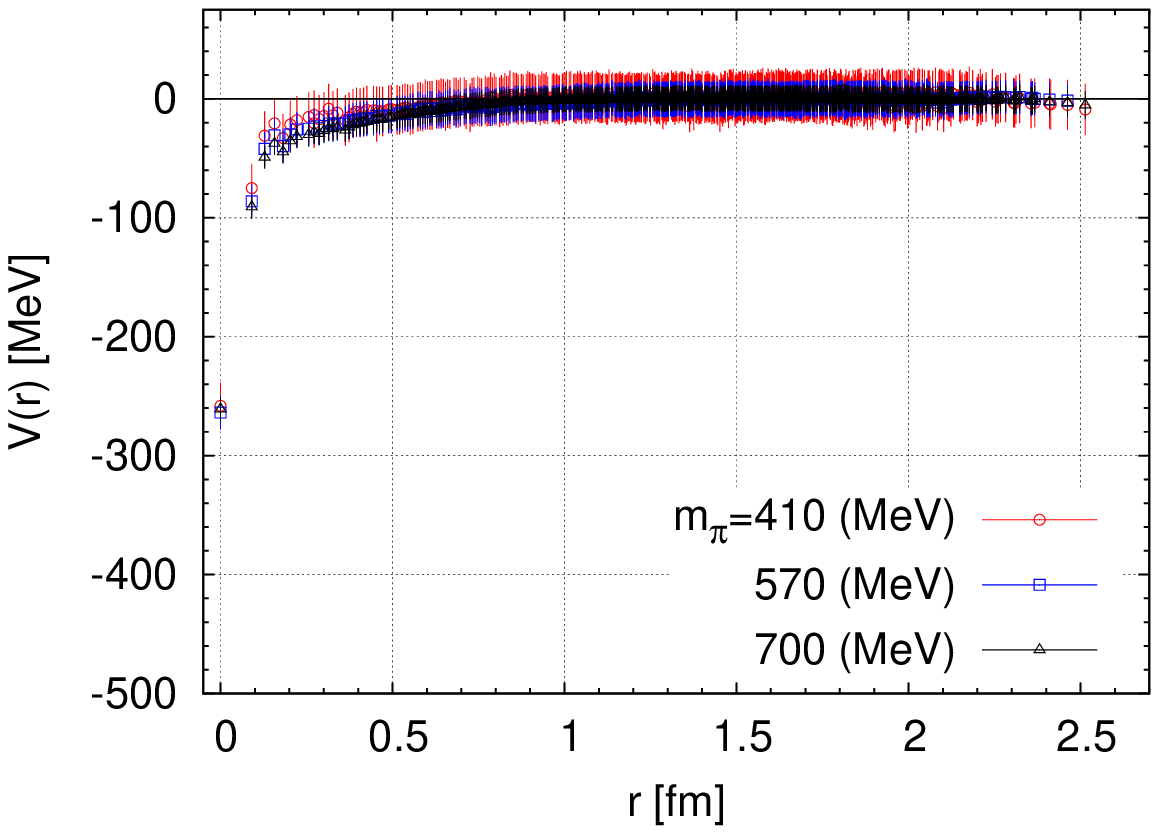}
\put(-80,50){\small (b) $V^{\rho \eta_c, \rho \eta_c}$}
\includegraphics[width=0.32\textwidth,clip]{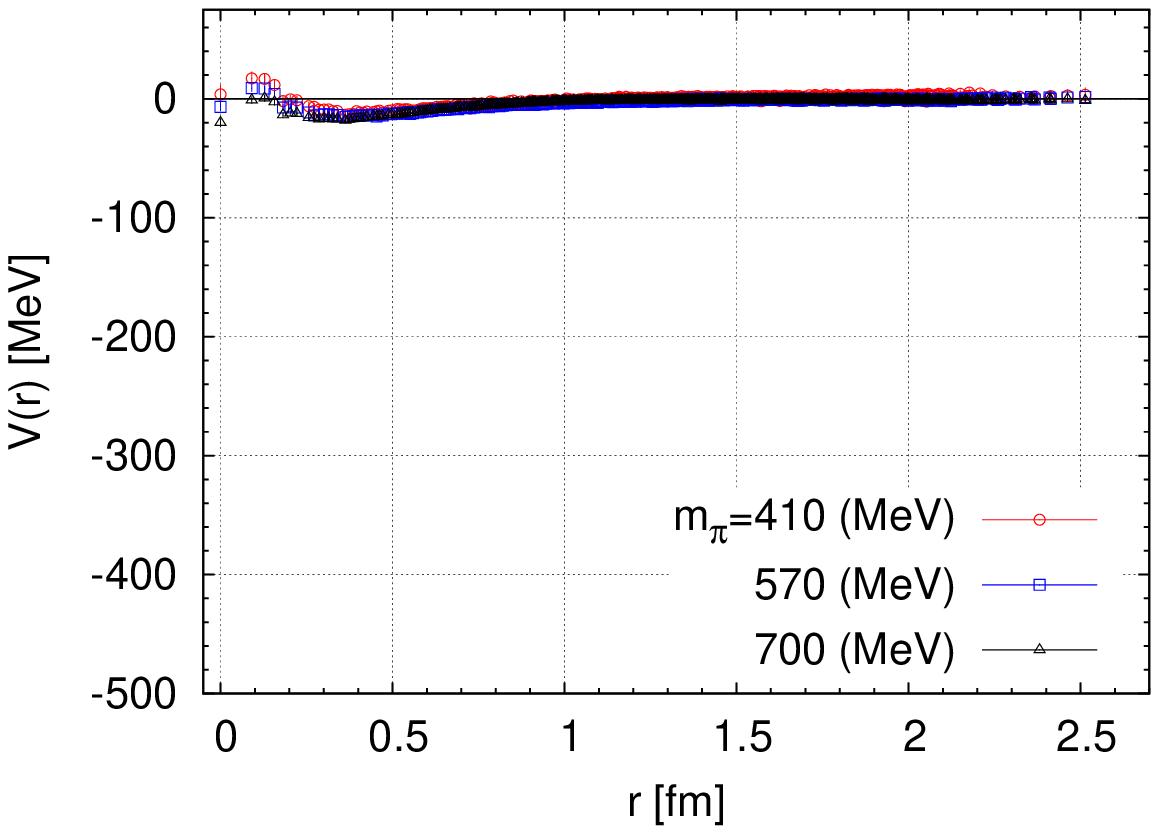}
\put(-80,50){\small (c) $V^{D \bar{D}^{*}, D \bar{D}^{*}}$}
\vspace{0.3cm}\\
\includegraphics[width=0.32\textwidth,clip]{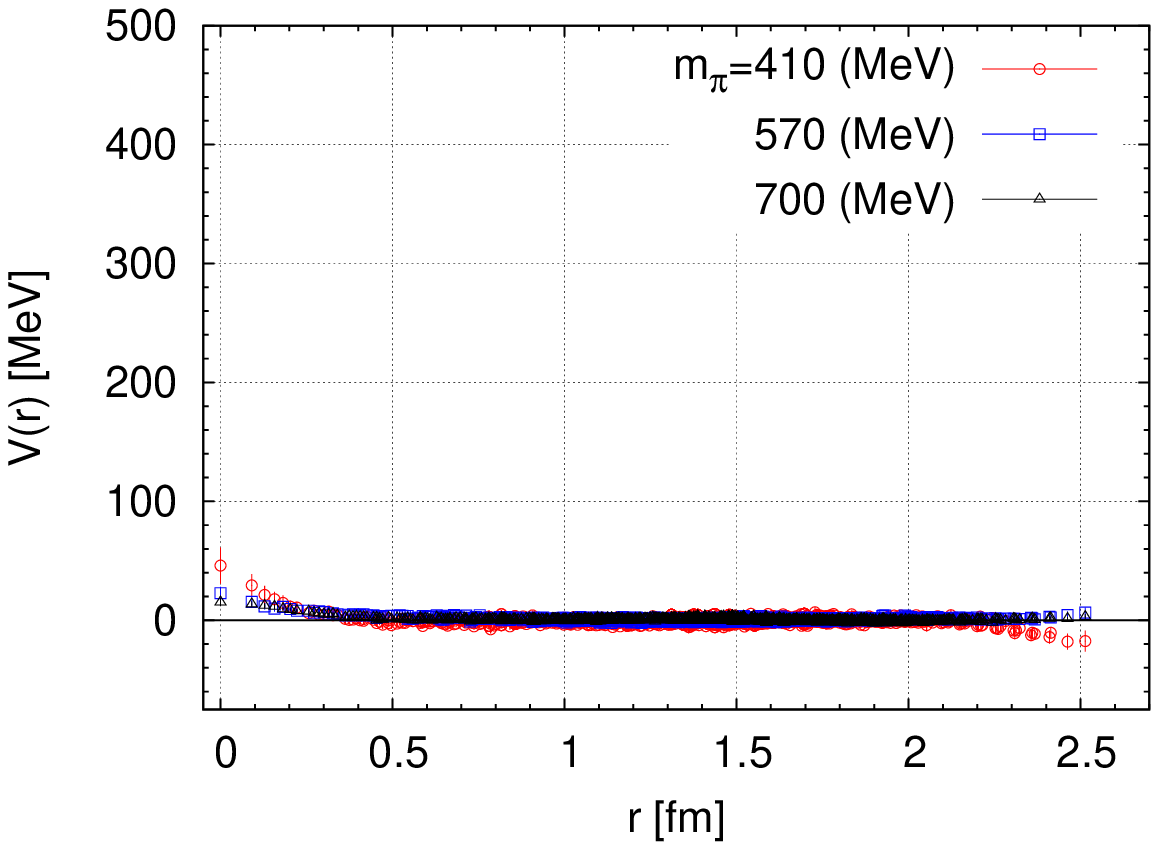}
\put(-80,50){\small (d) $V^{\pi J/\psi, \rho \eta_c}$}
\includegraphics[width=0.32\textwidth,clip]{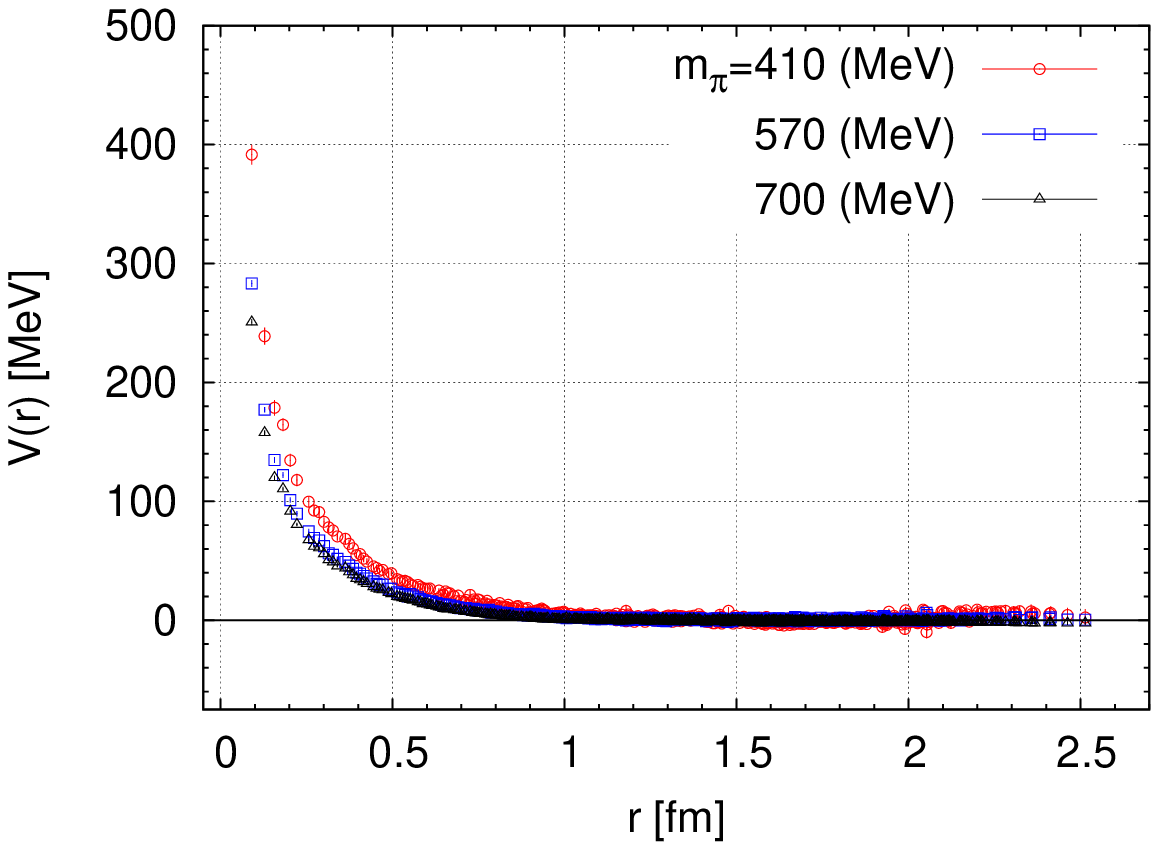}
\put(-80,50){\small (e) $V^{\pi J/\psi, D \bar{D}^{*}}$}
\includegraphics[width=0.32\textwidth,clip]{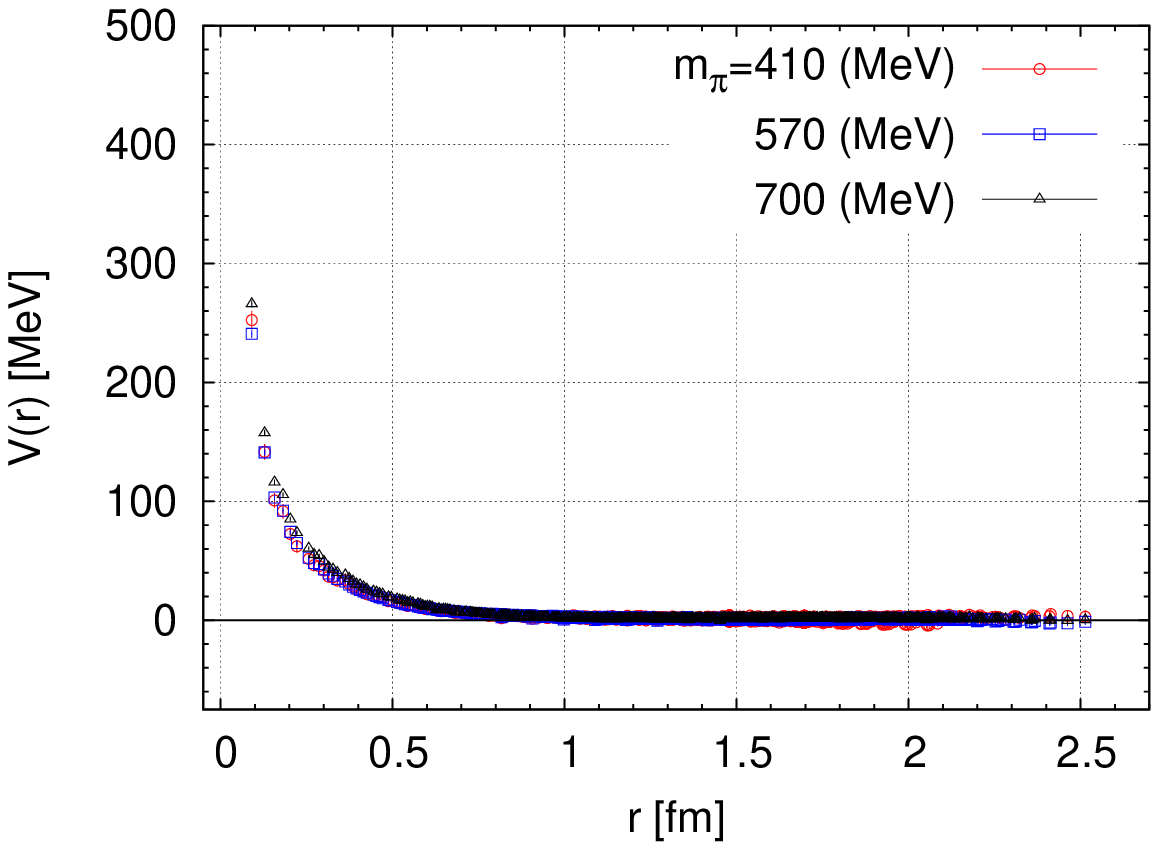}
\put(-80,50){\small (f) $V^{\rho \eta_c, D \bar{D}^{*}}$}
\caption{
The s-wave coupled-channel potential for the (a) $\pi J/\psi$-$\pi J/\psi$, (b) $\rho \eta_c$-$\rho \eta_c$, (c) $D \bar{D}^{*}$-$D \bar{D}^{*}$, (d) $\pi J/\psi$-$\rho \eta_c$, (e) $\pi J/\psi$-$D \bar{D}^{*}$ and (f) $\rho \eta_c$-$D \bar{D}^{*}$ channels.
The coupled-channel potentials are obtained at time slice $(t-t_{0})/a=13$
for case I(red circles), case II(blue squares) and case III(black triangles).
}
\label{fig2}
\end{figure*}

Shown in Fig.~\ref{fig2} is the resulting s-wave coupled-channel potential $V^{\alpha \beta}( r )$ from Eq.~(\ref{eq:s-wave_V}) at time slice $(t-t_0)/a=13$.
We find that all diagonal elements of the potential, 
(a) $V^{\pi J/\psi, \pi J/\psi}$, (b) $V^{\rho \eta_c, \rho \eta_c}$ 
and (c) $V^{D \bar{D}^{*}, D \bar{D}^{*}}$, are weak.
This observation indicates that the structure of the $Z_c(3900)$ is 
neither a simple $\pi J/\psi$ nor $D \bar{D}^{*}$ molecule.
Among the off-diagonal elements of the potential,
we find that the $\pi J/\psi$-$\rho \eta_c$ coupling in Fig.~\ref{fig2} (d) is also weak:
this is explained by the heavy-quark spin symmetry, 
which tells us that the spin flip amplitudes of the charm quark are suppressed by ${\cal O}(1/m_c)$.
On the other hand, the couplings,
(e) the $\pi J/\psi$-$D \bar{D}^{*}$ and (f) the $\rho \eta_c$-$D \bar{D}^{*}$, are both non-trivially strong:
they correspond to the rearrangement of quarks between the hidden charm sector and the open charm sector.
We will show later that these off-diagonal potentials play an important role 
to understand the structure of the $Z_c(3900)$.

\begin{figure*}[!htb]
\includegraphics[width=0.32\textwidth,clip]{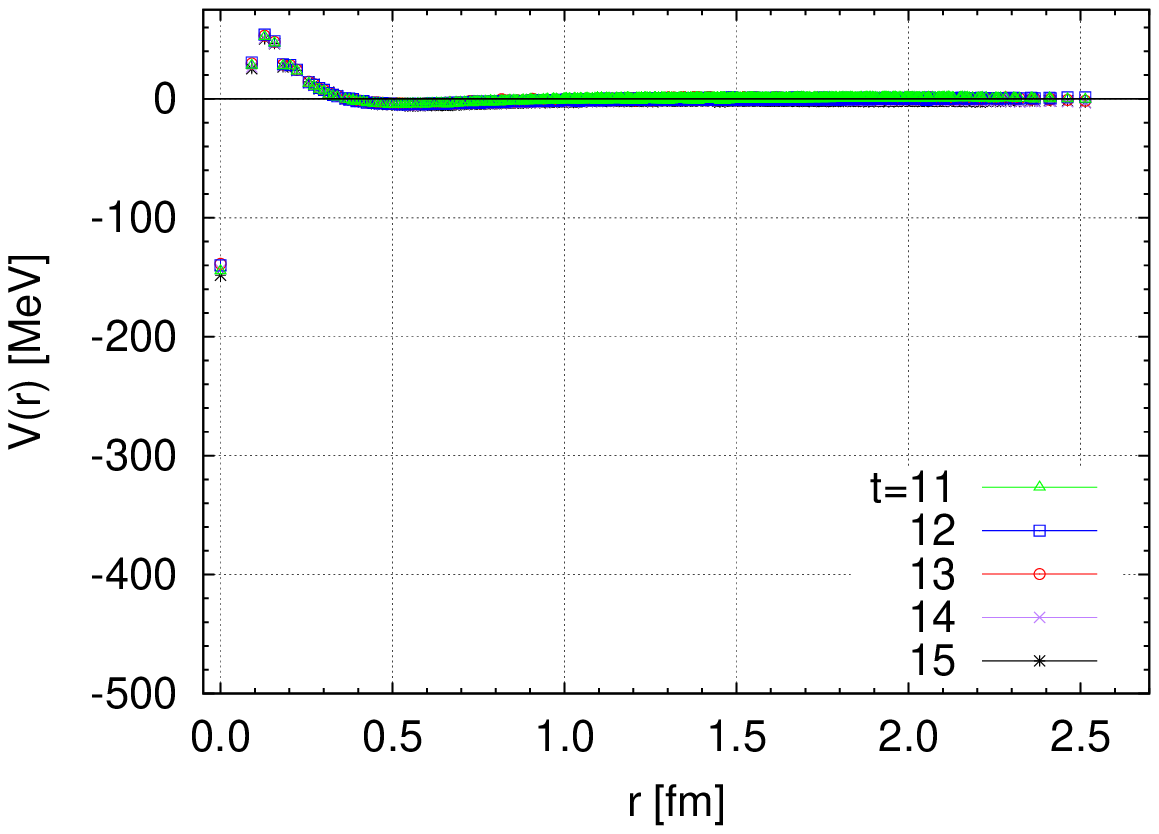}
\put(-80,50){\small (a) $V^{\pi J/\psi, \pi J/\psi}$}
\includegraphics[width=0.32\textwidth,clip]{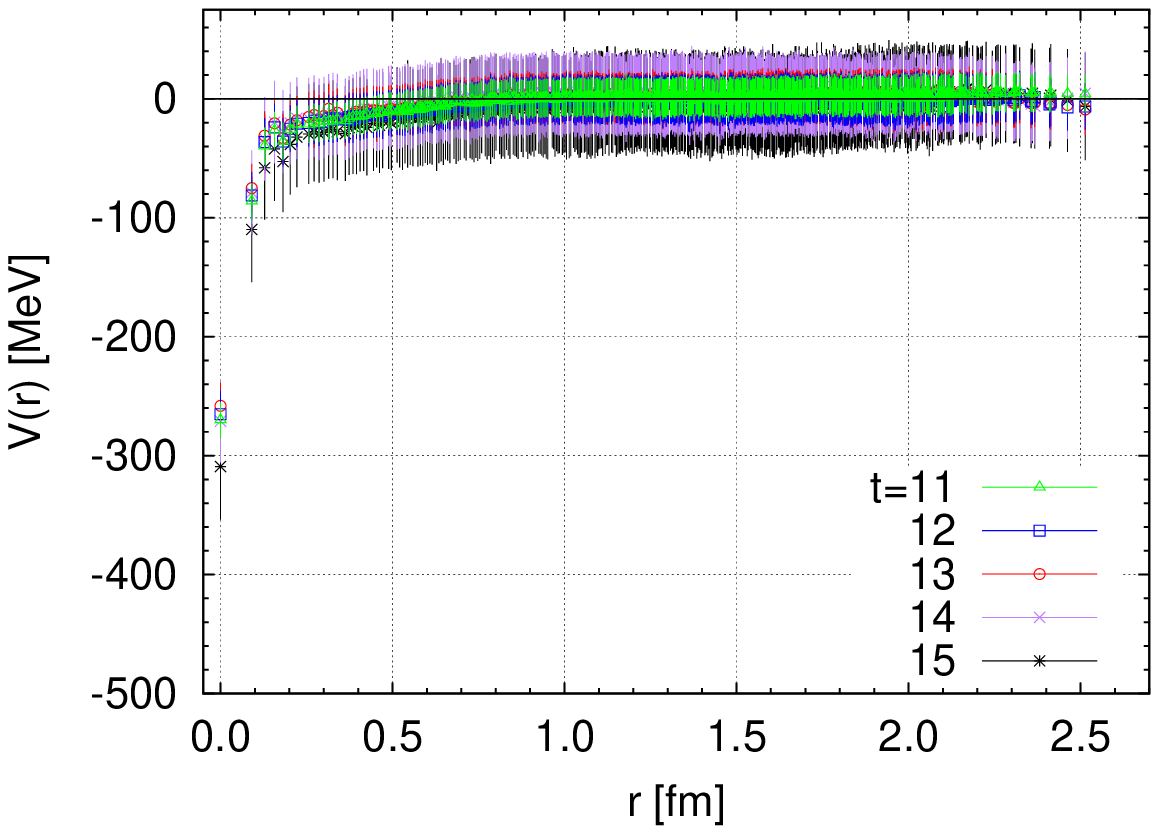}
\put(-80,50){\small (b) $V^{\rho \eta_c, \rho \eta_c}$}
\includegraphics[width=0.32\textwidth,clip]{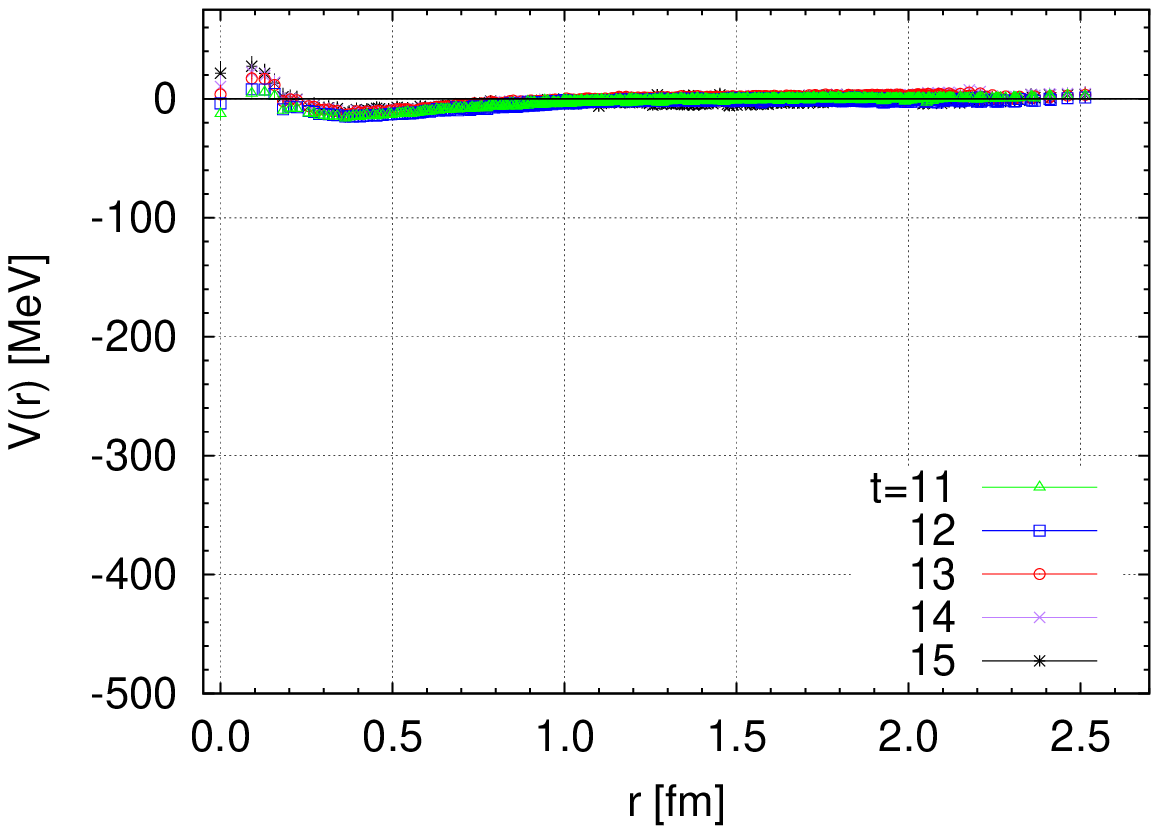}
\put(-80,50){\small (c) $V^{D \bar{D}^{*}, D \bar{D}^{*}}$}
\vspace{0.3cm}\\
\includegraphics[width=0.32\textwidth,clip]{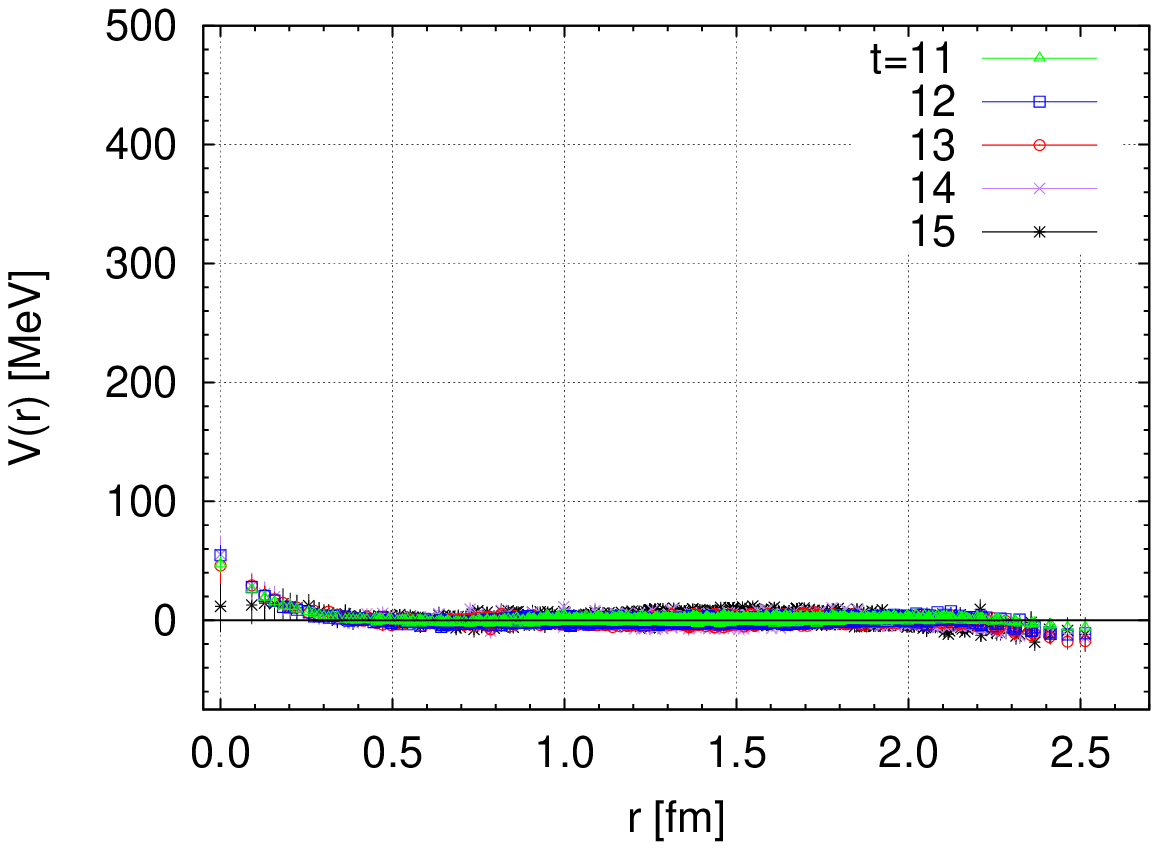}
\put(-80,50){\small (d) $V^{\pi J/\psi, \rho \eta_c}$}
\includegraphics[width=0.32\textwidth,clip]{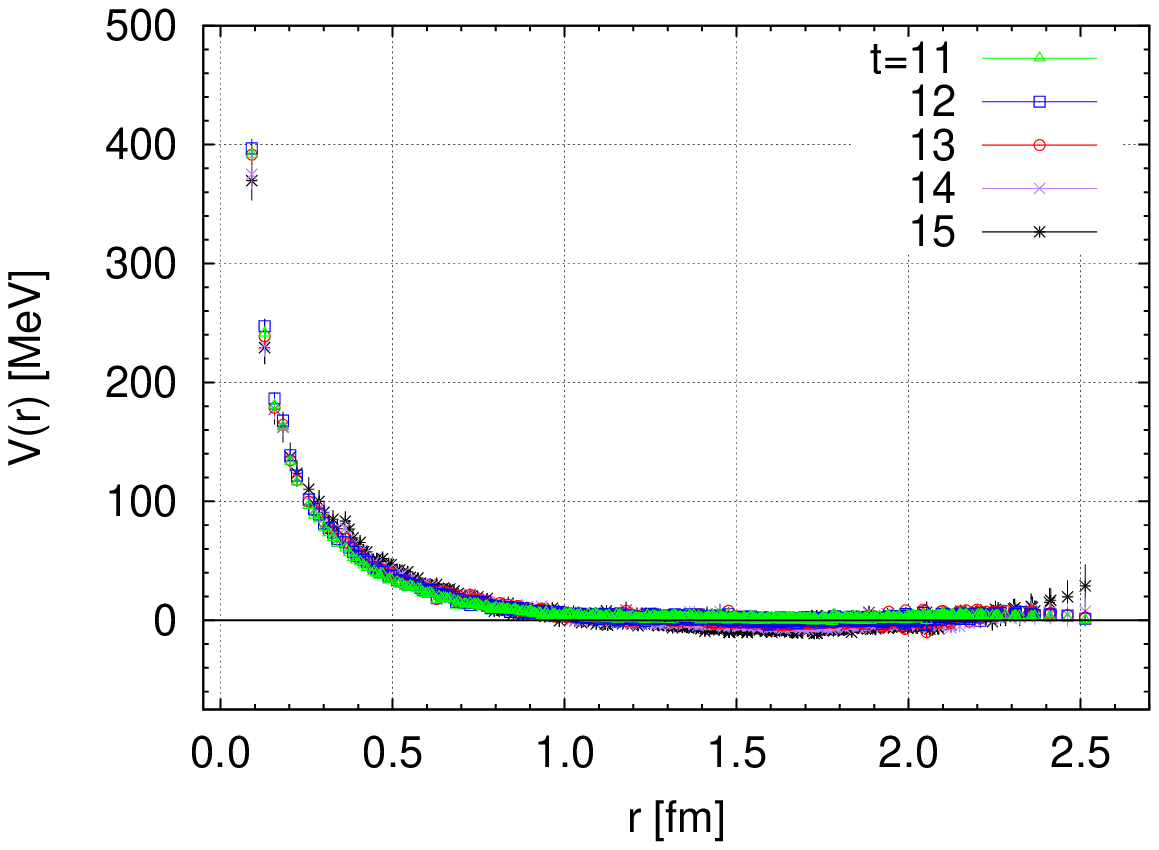}
\put(-80,50){\small (e) $V^{\pi J/\psi, D \bar{D}^{*}}$}
\includegraphics[width=0.32\textwidth,clip]{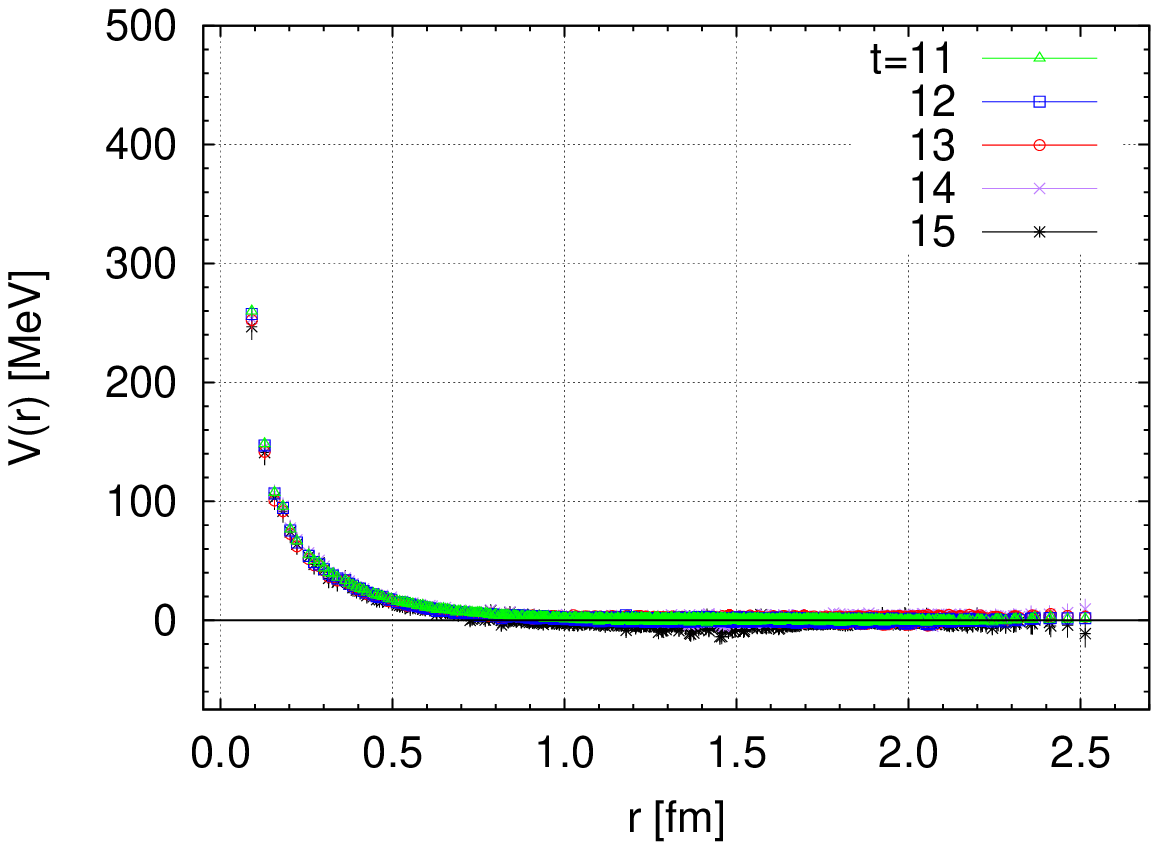}
\put(-80,50){\small (f) $V^{\rho \eta_c, D \bar{D}^{*}}$}
\caption{
Time slice dependence on the s-wave coupled-channel potential 
in $(t-t_0)/a=11$--$15$ for case I,
(a) $V^{\pi J/\psi, \pi J/\psi}$, (b) $V^{\rho \eta_c, \rho \eta_c}$, 
(c) $V^{D \bar{D}^{*}, D \bar{D}^{*}}$, (d) $V^{\pi J/\psi, \rho \eta_c}$, 
(e) $V^{\pi J/\psi, D \bar{D}^{*}}$ and (f) $V^{\rho \eta_c, D \bar{D}^{*}}$.
Similar time slice dependence to case II and III is also found.
}
\label{fig3}
\end{figure*}

We observe little time-slice dependence in $(t-t_0)/a=11$--$15$ on the potential $V^{\alpha \beta}$, as shown in Fig.~\ref{fig3}:
this implies that contributions from the inelastic $D^{*} \bar{D}^{*}$ scattering states to $V^{\alpha \beta}$ are negligible, and the convergence of the derivative expansion is reliable.

\begin{figure*}[!hbt]
\begin{center}
\includegraphics[width=0.32\textwidth,clip]{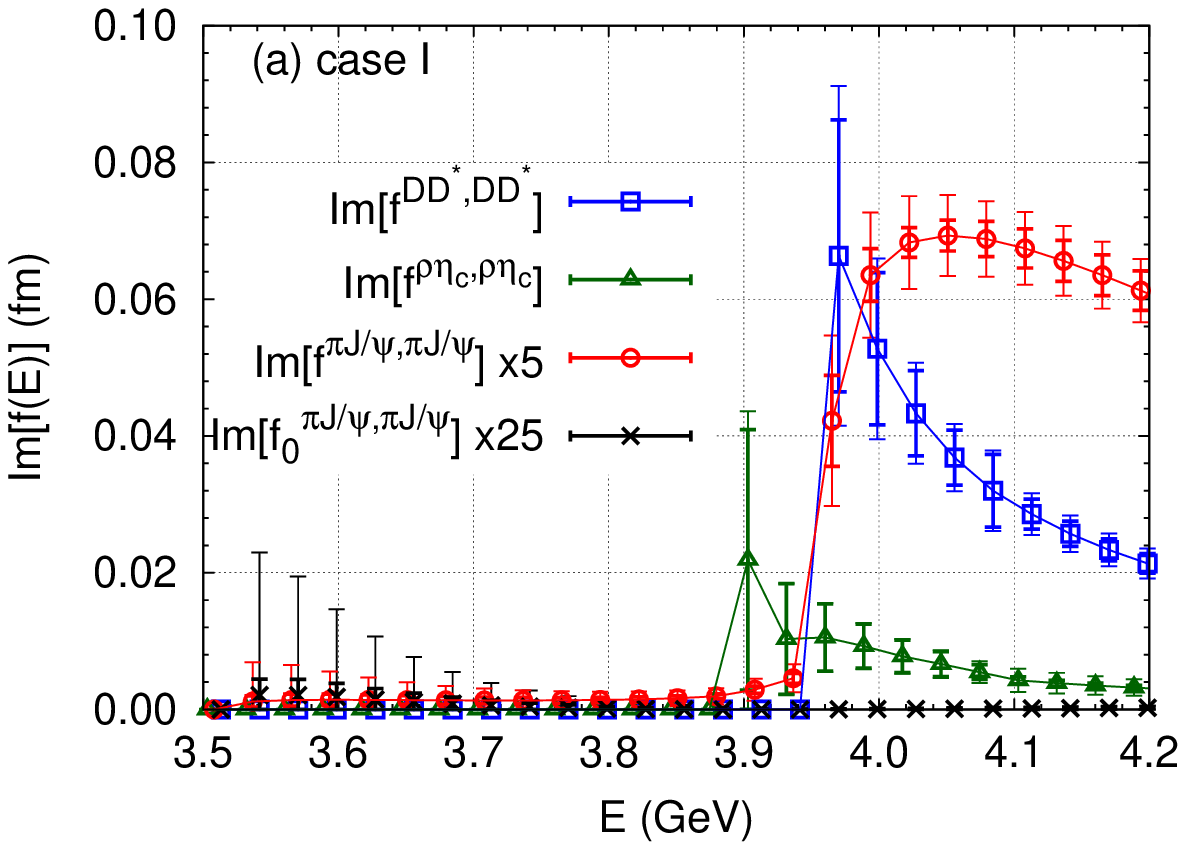}
\includegraphics[width=0.32\textwidth,clip]{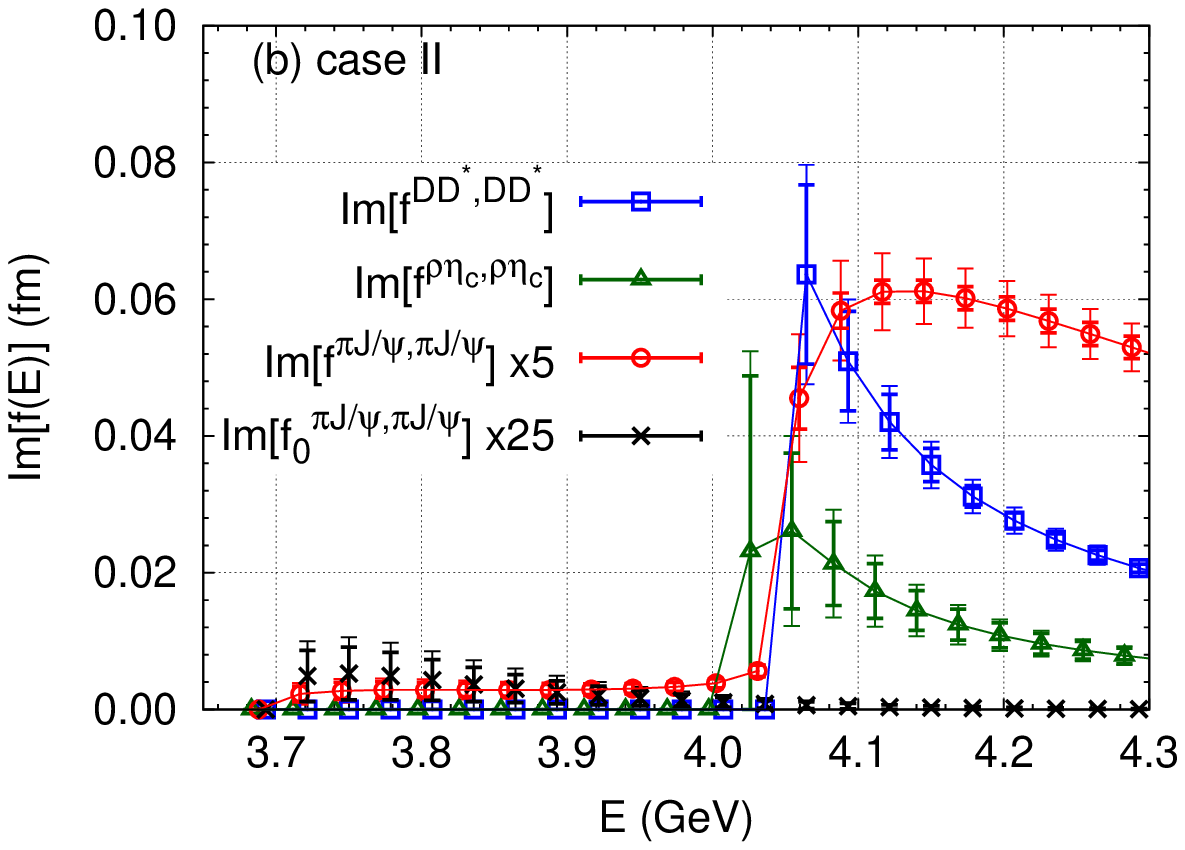}
\includegraphics[width=0.32\textwidth,clip]{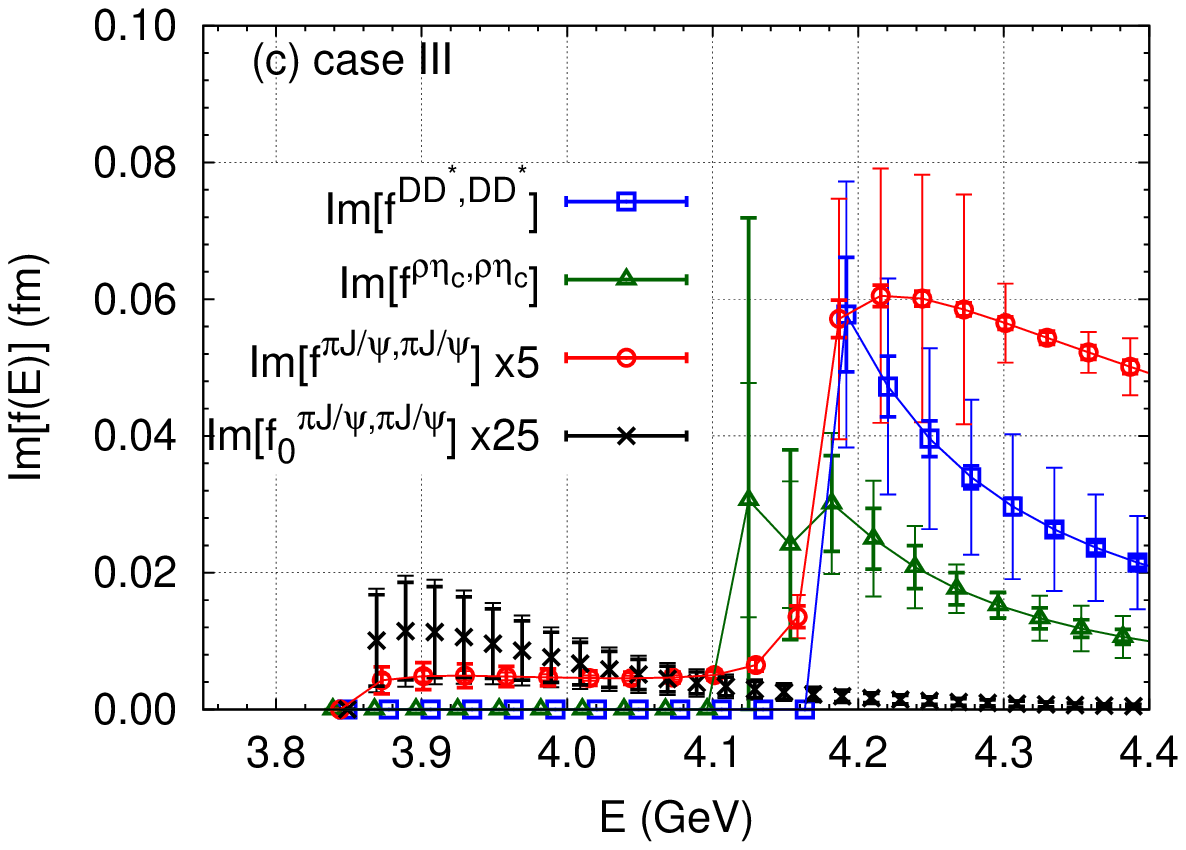}
\end{center}
\caption{
The two-body invariant mass spectra 
in the $\pi J/\psi$ (red circles, scaled by 5), 
$\rho \eta_{c}$ (green triangles) 
and $D \bar{D}^{*}$ (blue squares) channels 
for (a) case I, (b) case II and (c) case III in Table~\ref{tab1}.
The two-body $\pi J/\psi$ spectrum without the off-diagonal component of $V^{\alpha \beta}$ 
is also shown by $\mathrm{Im} f_{0}^{\pi J/\psi, \pi J/\psi}$ (black crosses, scaled by 25).
The inner error is statistical, while the outer one is statistical and systematic combined in quadrature.
}
\label{fig4}
\end{figure*}

As a first step to investigate the structure of the $Z_c(3900)$ on the basis of $V^{\alpha \beta}$ just obtained, 
we fit a three-range gauss function $f(r) = \sum_{n=1}^{3} g_{n} \exp[-a_n r^{2}]$ to the coupled-channel potential data.
Using the fit function, we obtain the coupled-channel potential 
in the momentum representation ($p$-space) through the Fourier transformation:
\begin{eqnarray}
V^{\alpha \beta}(\vec{p}_{\alpha}, \vec{p}_{\beta})
=
\int \frac{d\vec{r}}{(2\pi)^3} e^{ -i ( \vec{p}_{\alpha} - \vec{p}_{\beta} ) \cdot \vec{r} }f(r),
\label{eq:cc_pot.Fourier}
\end{eqnarray}
with $\vec{p}_{\alpha}$ being the relative momentum of the two-meson state in channel $\alpha$.
Since the potential $V^{\alpha \beta}(\vec{p}_{\alpha}, \vec{p}_{\beta})$ is the input to
the Lippmann-Schwinger (LS) equation,
let us consider the two-body T-matrix~\cite{Newton:book}: 
\begin{eqnarray}
t^{\alpha \beta}(\vec{p}_{\alpha}, \vec{p}_{\beta}; E)
& = &
V^{\alpha \beta}(\vec{p}_{\alpha}, \vec{p}_{\beta})
+ \sum_{\gamma} \int d\vec{q_{\gamma}}
\frac{ V^{\alpha \gamma}(\vec{p}_{\alpha}, \vec{q}_{\gamma}) 
t^{\gamma \beta}(\vec{q}_{\gamma}, \vec{p}_{\beta}; E)
}{
E - E_{\gamma}(\vec{q}_{\gamma}) + i \epsilon
}  ,
\label{eq:2-body.LS_eq}
\end{eqnarray}
where $\vec{p}_{\alpha}$ ($\vec{q}_{\gamma}$) indicates the on-shell (off-shell) momentum of the two-meson state in channel $\alpha$ ($\gamma$).
$E$ and $E_{\gamma}(\vec{q}_{\gamma})$ represent the scattering energy in the c.m. frame and the energy of the intermediate states in channel $\gamma$, respectively.
The T-matrix is related to the S-matrix and 
the scattering amplitude $f^{\alpha \beta}(E)$ as follows:
\begin{eqnarray}
\fl
S^{\alpha \beta}(\vec{p}_{\alpha}, \vec{p}_{\beta}; E) 
& = &
\delta_{\alpha \beta} \delta(\vec{p}_{\alpha} - \vec{p}_{\beta})
- 2 \pi i 
\delta\left(
E_{\alpha}(\vec{p}_{\alpha}) - E_{\alpha}(\vec{p}_{\beta})
\right)
t^{\alpha \beta}(\vec{p}_{\alpha}, \vec{p}_{\beta}; E)  ~,
\label{eq:2-body.S-mat}  
\\
\fl
f^{\alpha \beta}(\vec{p}_{\alpha}, \vec{p}_{\beta}; E)
& = &
- \pi \sqrt{ \mu^{\alpha} \mu^{\beta} } 
t^{\alpha \beta}(\vec{p}_{\alpha}, \vec{p}_{\beta}; E)  ~.
\label{eq:2-body.amp} 
\end{eqnarray}

Shown in Fig.~\ref{fig4} are the invariant mass spectra of the two-body scattering
which is the most ideal reaction process to understand the structure of the $Z_{c}(3900)$.
Since the s-wave amplitude $f^{\alpha \beta}(E)$ is related to the cross section as 
$\sigma^{\alpha \beta}(E) = 4 \pi |f^{\alpha \beta}(E)|^2$,
the two-body invariant mass spectra are given by $\mathrm{Im} f^{\alpha \alpha}(E)$ 
in the $\pi J/\psi$ (red circles), $\rho \eta_c$ (green triangles) 
and $D \bar{D}^{*}$ (blue squares) channels 
for (a) case I, (b) case II and (c) case III in Table \ref{tab1}.
In Fig.~\ref{fig4}, the inner errors are statistical only, while the outer ones are statistical and systematic errors added in quadrature:
the systematic errors from the truncation of the derivative expansion are evaluated by the difference between $\mathrm{Im} f^{\alpha \alpha}$ at $t=13$ and that at $t=15$.
The sudden enhancement in both the $\rho \eta_c$ and the $D \bar{D}^{*}$ spectra is caused by the opening of the s-wave thresholds.
The peak structure in the $\pi J/\psi$ spectrum just above the $D \bar{D}^{*}$ threshold is induced by the $\pi J/\psi$-$D \bar{D}^{*}$ coupling.
Indeed, if we switch off the off-diagonal components of $V^{\alpha \beta}$, 
the (red) circles turn into the (black) crosses without any peak structure at the $D \bar{D}^{*}$ threshold.
This fact implies that the peak structure in the $\pi J/\psi$ spectrum [called the $Z_c(3900)$]
is a typical ``threshold cusp''~\cite{Wigner:1948zz,Newton:book} due to the opening of the $s$-wave $D \bar{D}^{*}$ threshold.

\begin{figure*}[!h]
\begin{center}
\includegraphics[width=0.60\textwidth,clip]{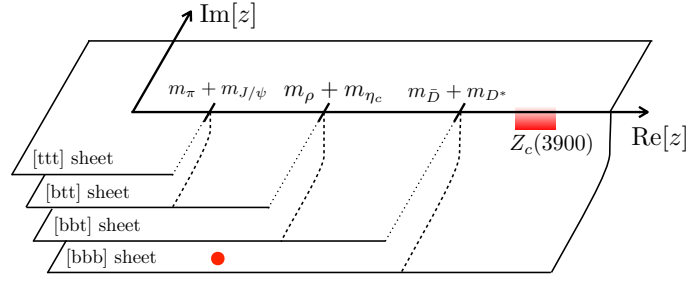}
\end{center}
\caption{
The complex energy plane in the $\pi J/\psi$-$\rho \eta_c$-$D \bar{D}^{*}$ coupled-channel system.
The energy relevant to $Z_c(3900)$ is indicated by shaded area.
Also, the pole position found in the numerical calculation is illustrated by red circle.
It is located far from the the physical region relevant to the  $Z_c(3900)$.
}
\label{fig5}
\end{figure*}

\begin{figure*}[!htb]
\begin{center}
\includegraphics[width=0.48\textwidth,clip]{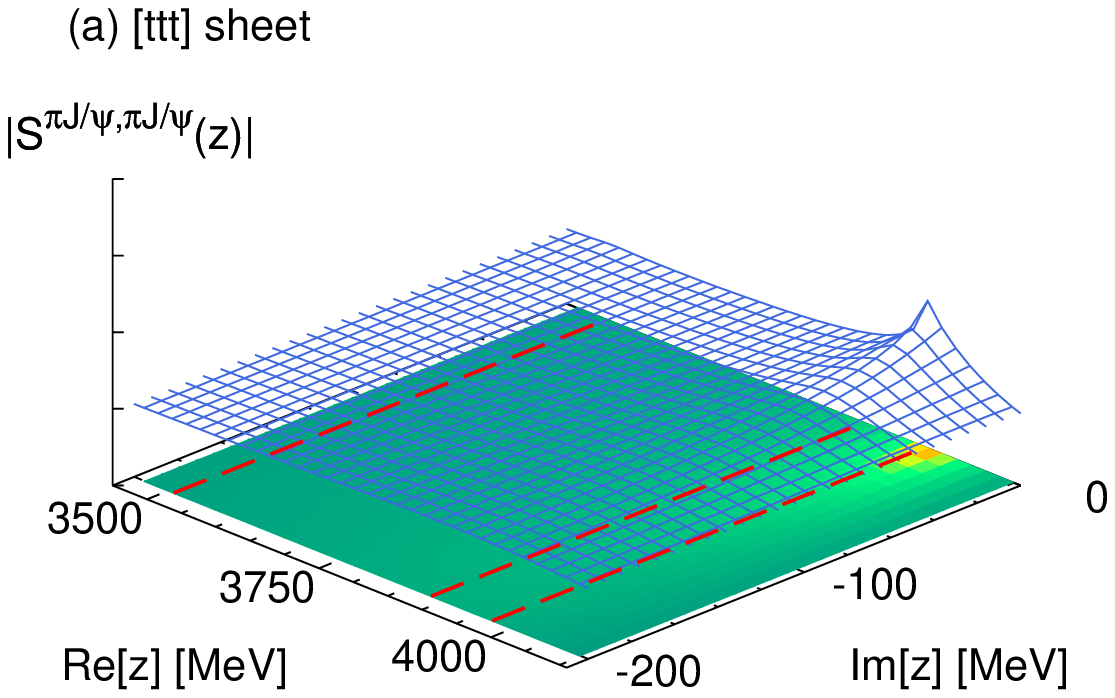}
\includegraphics[width=0.48\textwidth,clip]{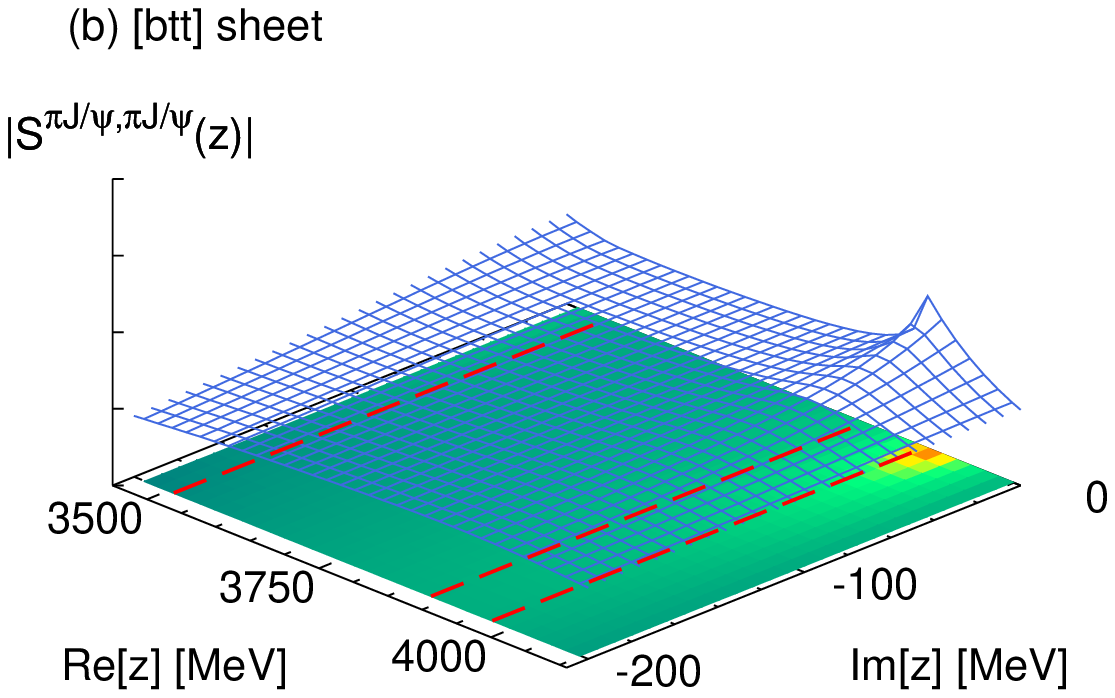}\\
\includegraphics[width=0.48\textwidth,clip]{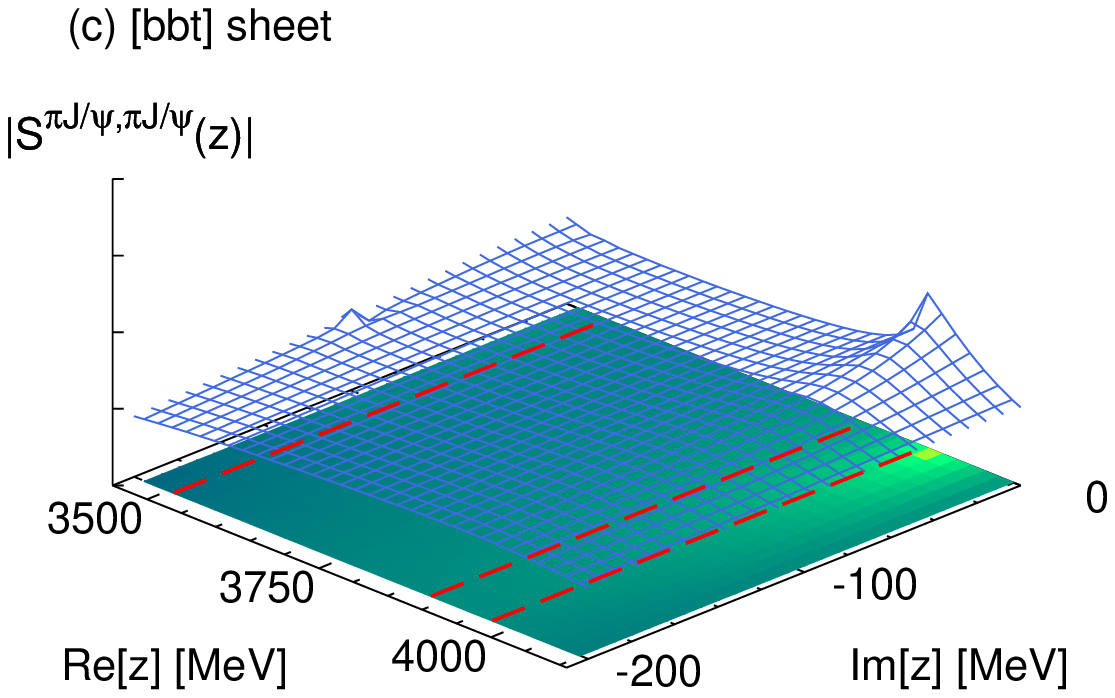}
\includegraphics[width=0.48\textwidth,clip]{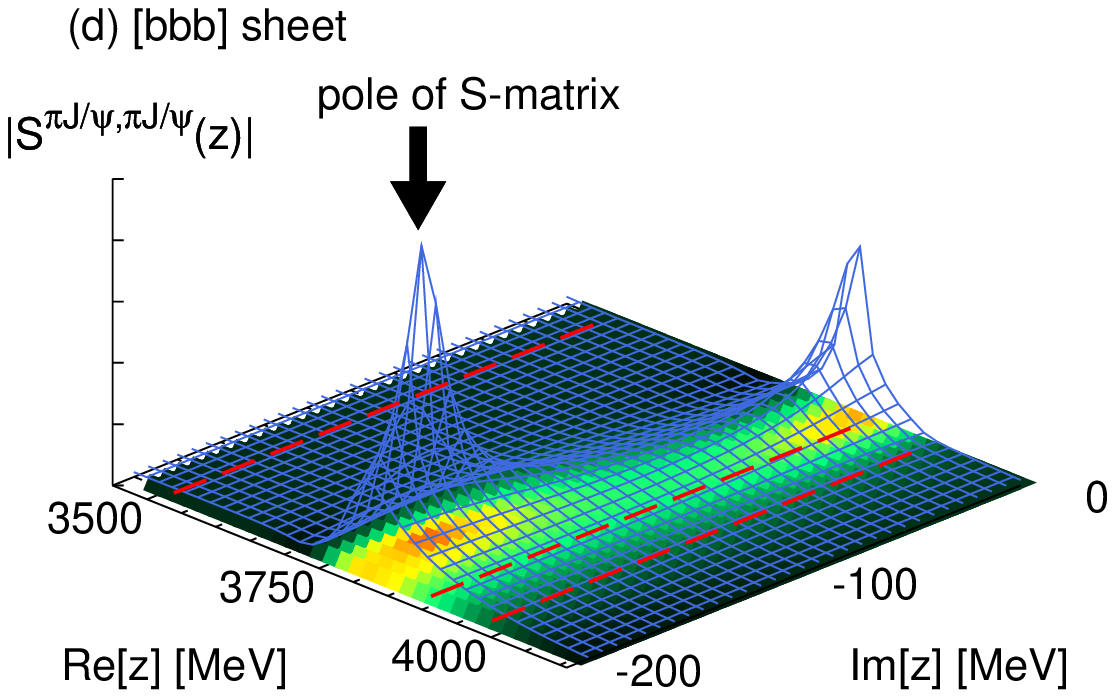}
\end{center}
\caption{
The absolute magnitude of the $\pi J/\psi$-$\pi J/\psi$ S-matrix on the 
(a) [$ttt$], (b) [$btt$], (c) [$bbt$] and (d) [$bbb$] Riemann sheets
in the notation of Ref.~\cite{Pearce:1988rk} 
for $\pi J/\psi$, $\rho \eta_c$ and $D \bar{D}^{*}$ channels.
The complex energy is defined by $z = m_1^{\alpha} + m_2^{\alpha} + p_{\alpha}^2/2\mu^{\alpha}$.
All figures correspond to case I in Table~\ref{tab1}.
The dashed lines represent the threshold energy $\mathrm{Re}[z] = m_1^{\alpha} + m_2^{\alpha}$.
The pole is found only on the [$bbb$] sheet, but the location is far from the $D \bar{D}^{*}$ threshold.
}
\label{fig6}
\end{figure*}

To make sure that the $Z_c(3900)$ is not associated with a resonance structure 
but a threshold cusp, 
we examine the pole positions of the S-matrix on the complex energy plane
according to the notation and procedure in Ref.~\cite{Pearce:1988rk} (see also Ref~\cite{Ikeda:2016zwx} and \ref{app:pole}).
The complex energy is defined by
$z = m_1^{\alpha} + m_2^{\alpha} + p_{\alpha}^2/2\mu^{\alpha}$, 
and the ``top [$t$]"  (``bottom [$b$]") sheet corresponds to 
$0 \le \arg p_{\alpha} < \pi$ ($\pi \le \arg p_{\alpha} < 2\pi$)
for the complex momentum in each channel ($\alpha = \pi J/\psi, \rho\eta_c, D \bar{D}^{*}$).
Among eight Riemann sheets for the present three-channel scattering, 
as shown in Fig.~\ref{fig5},
relevant sheets to the scattering amplitudes are the [$ttt$], [$btt$], [$bbt$] and [$bbb$] sheet in the notation of Ref.~\cite{Pearce:1988rk}.
If the $Z_c(3900)$ is a conventional resonance, the corresponding pole could be found just above the $D \bar{D}^{*}$ threshold with a negative imaginary part on the [$bbb$] sheet.
Shown in Fig.~\ref{fig6} are the absolute value of the $\pi J/\psi$-$\pi J/\psi$ S-matrix
on the (a) [$ttt$], (b) [$btt$], (c) [$bbt$] and (d) [$bbb$] sheet for case I in Table~\ref{tab1}, 
and we find a pole with a large imaginary part only on [$bbb$] sheet 
(The pole location is schematically illustrated in Fig.~\ref{fig5}.).
The complex energy of the pole is 
$z_{\rm pole}-(m_{D}+m_{\bar{D}^{*}})=-167(94)(27)-i183(46)(19)$ MeV for case I, 
$-128(76)(33)-i157(32)(19)$ MeV for case II, and $-190(56)(42)-i44(27)(27)$ MeV for case III, 
with the first and second parentheses indicating the statistical and systematic errors, respectively. 
In all cases, the pole is located far from the $D \bar{D}^{*}$ threshold on the real axis, 
so that the scattering amplitude is hardly affected by the pole.
With all analyses of the two-body coupled-channel scattering,
we conclude that the $Z_c(3900)$ is not a conventional resonance but a threshold cusp.

\section{Three-body decay of Y(4260)}
\label{results_3-body}
\begin{figure*}[!htb]
\begin{center}
\includegraphics[width=0.85\textwidth,clip]{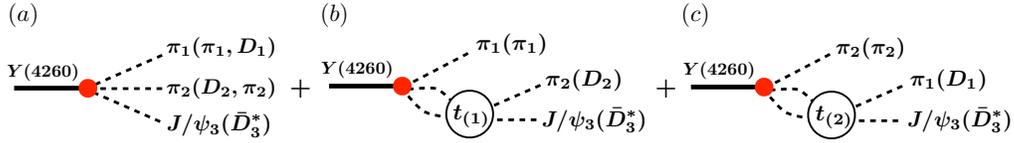}
\end{center}
\caption{
The diagrams and three-body Fock space 
of the $Y(4260) \to \pi \pi J/\psi$ ($\pi D \bar{D}^{*}$) decay:
(a) the background process, (b) and (c) the reaction processes.
}
\label{fig7}
\end{figure*}

In the previous section, 
it is found that the $Z_c(3900)$ is associated with the $D \bar{D}^{*}$ threshold cusp
induced by the off-diagonal components of $V^{\alpha \beta}$.
To make further connection between the result obtained from lattice QCD simulations 
and the experimentally observed structure in $\pi J/\psi$ and $D \bar{D}^{*}$ invariant mass spectra~\cite{expt_BESIII, Liu:2015jta, expt_Belle, expt_CLEO-c},
we analyze the three-body decays,
$Y(4260) \to \pi \pi J/\psi$ and $\pi D \bar{D}^{*}$ as shown in Fig.~\ref{fig7}, 
by taking into account the final state rescattering 
due to the coupled-channel potential extracted from lattice QCD simulations.

We define the three-body Fock space for the final states 
symmetric with respect to identical two pions as follows:
\begin{eqnarray}
\fl
\left|
\pi(\vec{p}_i) \otimes \left[  A(\vec{p}_j) \otimes B(\vec{p}_k)  \right]_{\alpha}
\right\rangle 
=
\left\{ \begin{array}{ll}
\left| 
\pi_{1}(\vec{p}_{1}) ~ \pi_{2}(\vec{p}_{2}) ~ J/\psi_{3}(\vec{p}_{3}) ; \alpha
\right\rangle
\\
\frac{1}{\sqrt{2}}
\left(
\left| 
\pi_{1}(\vec{p}_{1}) ~ D_{2}(\vec{p}_{2}) ~ \bar{D}^{*}_{3}(\vec{p}_{3}) ; \alpha
\right\rangle 
+
\left| 
D_{1}(\vec{p}_{1}) ~ \pi_{2}(\vec{p}_{2}) ~ \bar{D}^{*}_{3}(\vec{p}_{3}) ; \alpha
\right\rangle 
\right)
\\
\end{array} \right.   , \nonumber \\
\label{eq:3-body_Fock}
\end{eqnarray}
Here $[A \otimes B]_{\alpha}=[\pi \otimes J/\psi]$, $[D \otimes \bar{D}^{*}]$ represents 
the interacting pair in the final state rescattering, 
whose flavor channels are specified by $\alpha$;
the indices $i,j,k=1,2,3$ label the spectator particle with momentum $\vec{p}_{i,j,k}$
in the three-body channel $\pi + \alpha$ $(\alpha = \pi J/\psi, D \bar{D}^{*})$; 
all the momenta are defined in the $Y(4260)$ rest frame ($\vec{p}_k = -\vec{p}_i-\vec{p}_j$).
In what follows, we denote the three-body final states in Eq.~(\ref{eq:3-body_Fock})
as $| \vec{p}_{i}, \vec{p}_{j}, \vec{p}_{k}; \alpha \rangle$ for simplicity.
The three-body state is normalized as
$\langle \vec{p'}_{i}, \vec{p'}_{j}, \vec{p'}_{k}; \alpha | \vec{p}_{i}, \vec{p}_{j}, \vec{p}_{k}; \beta \rangle = \delta_{\alpha \beta} \delta(\vec{p'}_{i} - \vec{p}_{i}) \delta(\vec{p'}_{j} - \vec{p}_{j})$.

The three-body amplitude relevant to the $Y(4260) \to \pi + \alpha$ decay is given by
\begin{eqnarray}
\fl
\left\langle Y(4260) \right| 
T_{Y \to \pi+\alpha}(W_3) 
\left| \vec{p}_{i}, \vec{p}_{j}, \vec{p}_{k}; \alpha \right\rangle
\nonumber \\
= 
\left\langle Y(4260) \right|
\left\{
\Gamma + \Gamma G_{0}(W_3) \sum_{i=1,2} t_{(i)}( W_3-E_{\pi}(\vec{p}_{i}), \vec{p}_{i} )
\right\}
\left| \vec{p}_{i}, \vec{p}_{j}, \vec{p}_{k}; \alpha \right\rangle ,
\end{eqnarray}
where $|Y(4260) \rangle$ denotes the $Y(4260)$ state;
the mass of the $Y(4260)$ and the energy of the spectator pion
are denoted by $W_3$ and $E_{\pi}(\vec{p}_{i})$, respectively.
The operator $\Gamma$ represents the primary vertex of the $Y(4260)$ decay.
Three-body free Green's function $G_0(W_3)$
is given by
\begin{eqnarray}
\fl
\langle \vec{p'}_{i}, \vec{p'}_{j}, \vec{p'}_{k}; \alpha | 
G_{0}(W_3) 
| \vec{p}_{i}, \vec{p}_{j}, \vec{p}_{k}; \beta \rangle
& = &
\frac{
\delta_{\alpha \beta}
\delta(\vec{p'}_{i} - \vec{p}_{i})
\delta(\vec{p'}_{j} - \vec{p}_{j})
}{
W_3 - E_{i}(\vec{p}_{i}) - E_{j}(\vec{p}_{j}) - E_{k}(\vec{p}_{k}) + i \epsilon
}  ~,
\end{eqnarray}
with
$E_{i}(\vec{p}_{i}) = \sqrt{m_{i}^{2} + \vec{p}_{i}^{2}}$.
To derive the two-body t-matrix $t_{(i)}(W_3-E_{\pi}(\vec{p}_{i}), \vec{p}_{i})$
in the presence of a spectator pion $\pi_{i}$ ($i=1,2$) in Eq.~(\ref{eq:3-body_Fock}),
it is useful to introduce the relative momentum $\vec{q}_{\alpha}$ 
for an interacting $jk$-pair in its c.m. frame,
then the momenta $\vec{p}_{j}$ and $\vec{p}_{k}$ are related to 
the momenta $\vec{p}_{i}$ and $\vec{q}_{\alpha}$ by a Lorentz boost,
\begin{eqnarray}
\vec{p}_j 
& = &
\vec{q}_{\alpha} - \frac{\vec{p}_i}{M_{jk}(\vec{q}_{\alpha})}
\left[
E_j(\vec{q}_{\alpha}) - 
\frac{\vec{p}_i \cdot \vec{q}_{\alpha}}{E_{jk}(\vec{p}_i, \vec{q}_{\alpha}) + M_{jk}(\vec{q}_{\alpha})}
\right] ~, \\
\vec{p}_k
& = &
-\vec{q}_{\alpha} - \frac{\vec{p}_i}{M_{jk}(\vec{q}_{\alpha})}
\left[
E_k(\vec{q}_{\alpha}) +
\frac{\vec{p}_i \cdot \vec{q}_{\alpha}}{E_{jk}(\vec{p}_i, \vec{q}_{\alpha}) + M_{jk}(\vec{q}_{\alpha})}
\right] ,
\end{eqnarray}
with 
the invariant mass of an interacting $jk$-pair, $M_{jk}(\vec{q}_{\alpha}) = E_{j}(\vec{q}_{\alpha}) + E_{k}(\vec{q}_{\alpha})$, and
the energy of the pair,
$E_{jk}(\vec{p}_i, \vec{q}_{\alpha}) = \sqrt{M_{jk}(\vec{q}_{\alpha})^{2} + \vec{p}_{i}^{2}}$.
Using the spectator pion momentum $\vec{p}_i$ 
and relative momentum $\vec{q}_{\alpha}$ in channel $\alpha$,
the two-body t-matrix is evaluated as follows:
\begin{eqnarray}
\fl
\langle \vec{p'}_{i}, \vec{p'}_{j}, \vec{p'}_{k}; \alpha | 
t_{(i)}(W_3-E_{\pi}(\vec{p}_{i}), \vec{p}_{i})
| \vec{p}_{i}, \vec{p}_{j}, \vec{p}_{k}; \beta \rangle
=
\delta(\vec{p'}_{i} - \vec{p}_{i})
t_{(i)}^{\alpha \beta}(\vec{q}_{\alpha}, \vec{q}_{\beta}, \vec{p}_{i}; W_3) ~,
\nonumber \\
\fl
t_{(i)}^{\alpha \beta}(\vec{q}_{\alpha}, \vec{q}_{\beta}, \vec{p}_{i}; W_3)
=
V_{(i)}^{\alpha \beta}(\vec{q}_{\alpha}, \vec{q}_{\beta}) 
+ 
\sum_{\gamma = \pi J/\psi, \rho\eta_c, D\bar{D}^{*}} 
\int d\vec{q}_{\gamma}
\frac{
V_{(i)}^{\alpha \gamma}(\vec{q}_{\alpha}, \vec{q}_{\gamma})
t_{(i)}^{\gamma \beta}(\vec{q}_{\gamma}, \vec{q}_{\beta}, \vec{p}_{i}; W_3)
}{
W_3 - E_{\pi}(\vec{p}_{i}) - E_{jk}(\vec{p}_{i},\vec{q}_{\gamma}) + i\epsilon 
}  ~, \nonumber \\
\end{eqnarray}
with 
\begin{eqnarray}
V_{(i)}^{\alpha \beta}(\vec{q}_{\alpha}, \vec{q}_{\beta}) 
=
\sqrt{ {\cal N}^{\alpha}(\vec{q}_{\alpha}) }
V^{\alpha \beta}(\vec{q}_{\alpha}, \vec{q}_{\beta}) 
\sqrt{ {\cal N}^{\beta}(\vec{q}_{\beta}) } ~,
\end{eqnarray}
where $V^{\alpha \beta}(\vec{q}_{\alpha}, \vec{q}_{\beta})$ is given in Eq.~(\ref{eq:cc_pot.Fourier}),
and
${\cal N}^{\alpha}(\vec{q}_{\alpha}) = \mu_i^{\alpha} / \mu_i^{\alpha}(\vec{q}_{\alpha})$
is a conventional factor to account for the relativistic kinematics;
$\mu_i^{\alpha} = m^{\alpha}_j m^{\alpha}_k/(m^{\alpha}_j + m^{\alpha}_k)$ and
$\mu_i^{\alpha}(\vec{q}_{\alpha}) = E_j(\vec{q}_{\alpha}) E_k(\vec{q}_{\alpha})/M_{jk}(\vec{q}_{\alpha})$
are the reduced mass and the reduced energy of particles $j$ and $k$, respectively.

Modeling the primary vertex by complex constants,
$C_{Y \rightarrow \pi + \alpha} \equiv \langle Y(4260)| \Gamma | \vec{p}_{i},\vec{p}_{j},\vec{p}_{k}; \alpha \rangle$ ($\alpha=\pi J/\psi, D \bar{D}^{*}$),
the three-body amplitudes for $T_{Y \rightarrow  \pi + \beta}$ ($\beta = \pi J/\psi, D \bar{D}^{*}$)
are now given by
\begin{eqnarray}
\fl
T_{Y\rightarrow \pi + \beta}(\vec{p}_i, \vec{q}_{\beta}; W_3)
= 
C_{Y \rightarrow \pi+\beta}
+
\sum_{\alpha= \pi J/\psi,  D \bar{D}^{*} } 
f_{\alpha} 
f_{\beta}
C_{Y \rightarrow \pi+\alpha}
\nonumber \\
\fl
~~ \times
\biggl( 
\int d\vec{q}_{\alpha} 
\frac{ 
t_{(i)}^{\alpha \beta}(\vec{q}_{\alpha}, \vec{q}_{\beta}, \vec{p}_i; W_3)
}{
W_3 - E_{i}(\vec{p}_{i}) - E_{j}(\vec{p}_{j}) - E_{k}(\vec{p}_{k}) + i\epsilon 
}
+
\int d\vec{q}_{\alpha (j)} 
\frac{ 
t_{(j)}^{\alpha \beta}
(\vec{q}_{\alpha (j)}, \vec{q}_{\beta (j)}, \vec{p}_j(\vec{p}_i, \vec{q}_{\beta}); W_3)
}{
W_3 - E_{i}(\vec{p}_{i}) - E_{j}(\vec{p}_{j}) - E_{k}(\vec{p}_{k}) + i\epsilon 
}
\biggr) ~,  \nonumber \\
\label{eq:3-body_amp}
\end{eqnarray}
with $(i,j)=(1,2)$ or $(2,1)$ and $(f_{\pi J/\psi}, f_{D \bar{D}^{*}}) = (1, 1/\sqrt{2})$.
The relative momentum of an interacting $ik$ pair is denoted by $\vec{q}_{\alpha (j)}$.
With the three-body amplitude, 
the decay rate $\Gamma_{Y \to \pi + \alpha}$ is give by
\begin{eqnarray}
\fl
\Gamma_{Y \to \pi + \alpha}(W_3)
& = & 
(2 \pi)^4 
\int d \vec{p}_i  d \vec{p}_{j}
\delta( W_3 - E_{i}(\vec{p}_i) - E_{j}(\vec{p}_j) - E_{k}(\vec{p}_k) )
\left| T_{Y\rightarrow \pi + \alpha}(\vec{p}_i, \vec{q}_{\alpha}; W_3) \right|^2 
\nonumber \\
\fl
& = & 
(2 \pi)^4 
\int dM_{\alpha}  d\hat{p}_{i} d\hat{q}_{\alpha}
\frac{ E_i(\vec{p}_i) E_j(\vec{p}_j) E_k(\vec{p}_k) }{W_3} p_{i} q_{\alpha}
\left| T_{Y\rightarrow \pi + \alpha}(\vec{p}_i, \vec{q}_{\alpha}; W_3) \right|^2 
~,
\end{eqnarray}
where $M_{\alpha} = M_{jk}(\vec{q}_{\alpha})$ is an invariant mass in channel $\alpha$.
Therefore, the invariant mass spectrum is found as
\begin{eqnarray}
\fl
\frac{ d\Gamma_{Y \to \pi + \alpha}(W_3) }{dM_{\alpha}}
=
(2 \pi)^4 8\pi^2 
\int d(\hat{p}_{i} \cdot \hat{q}_{\alpha})
\frac{ E_i(\vec{p}_i) E_j(\vec{p}_j) E_k(\vec{p}_k) }{W_3} p_{i} q_{\alpha}
\left| T_{Y\rightarrow \pi + \alpha}(\vec{p}_i, \vec{q}_{\alpha}; W_3) \right|^2 
~.
\label{eq:decay_inv_mass}
\end{eqnarray}
We note that in Eq.~(\ref{eq:decay_inv_mass}) all three particles are assumed to be in S-state.
It is also noted that, in Ref.~\cite{Ikeda:2016zwx}, we have employed the non-relativistic kinematics, but the relativistic one is used in the present study.

\begin{figure}[!htb]
\begin{center}
\includegraphics[width=0.48\textwidth,clip]{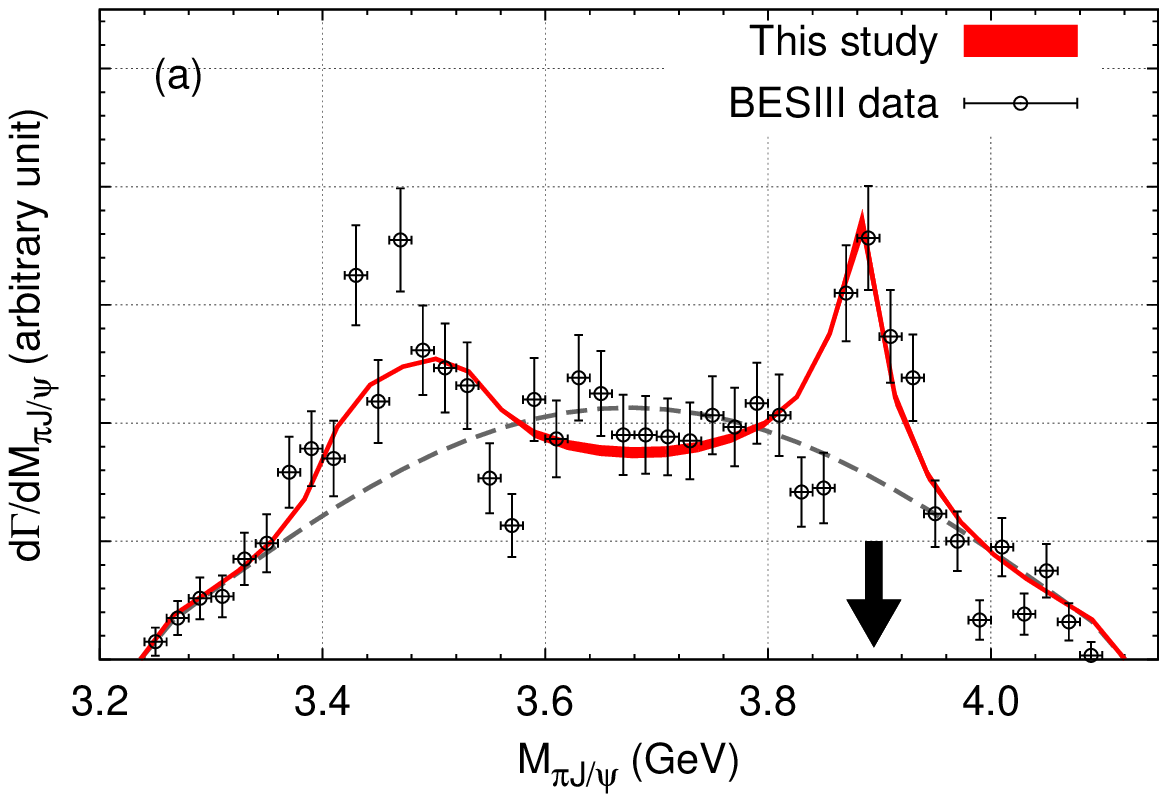}
\includegraphics[width=0.48\textwidth,clip]{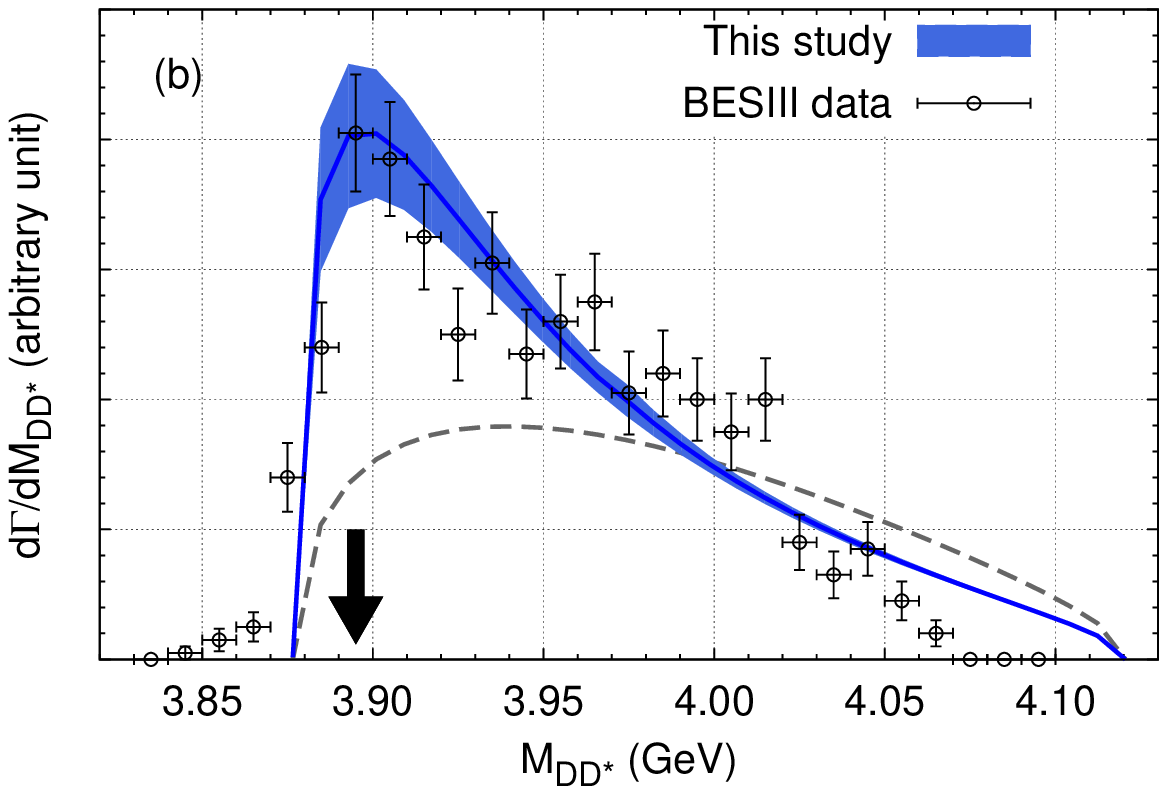}
\end{center}
\caption{
The invariant mass spectra of (a) $Y(4260) \to \pi \pi J/\psi$ and (b)  $Y(4260) \to \pi D \bar{D}^{*}$
calculated with $V^{\alpha \beta}$ for case I in Table \ref{tab1}.
The shaded areas show the statistical errors.
The vertical arrows show the predicted peak positions from the calculations.
The blue dashed lines show the invariant mass spectra of the $Y(4260)$ decay without the off-diagonal components of $V^{\alpha \beta}$.
The experimental data are taken from Ref.~\cite{expt_BESIII,Liu:2015jta}.
}
\label{fig8}
\end{figure}

In the analysis of the decay rate, 
we employ the physical hadron masses to have the same phase space as the experiments,
while $t_{(i,j)}^{\alpha \beta}$ is taken into account from the lattice data for case I.
The three-body energy is fixed at the $Y(4260)$ energy ($W_3=4260$ MeV), and 
the complex couplings $C_{Y\rightarrow \pi+\alpha}$ are fitted to 
the $\pi J/\psi$ and $D \bar{D}^{*}$ invariant mass spectra
obtained by BESIII Collaboration
shown in Fig.~\ref{fig8}~\cite{expt_BESIII,Liu:2015jta}.
For the $D \bar{D}^{*}$ invariant mass spectrum, 
recently updated double $D$-tag data is used~\cite{Liu:2015jta}.
Since the experimental data are in the arbitrary scale,
we first focus only on the line shapes of the invariant mass spectra.
In this case, we have two real parameters, 
$R \equiv | C_{Y \rightarrow \pi (D \bar{D}^{*})} / C_{Y \rightarrow \pi (\pi J/\psi)} |$ and $\theta \equiv \arg ( C_{Y \rightarrow \pi (D \bar{D}^{*})} / C_{Y \rightarrow \pi (\pi J/\psi)} )$, 
and the best fit values are $R=5.20(17)$ and $\theta=-11.4(3)$ degree, 
where we consider the statistical error only.
To compare with the raw data of the experiment,
we find 
$\mathcal{N}_{\pi J/\psi} |C_{Y \to \pi (\pi J/\psi)}|=1.59(78) \times 10^{-3}$ MeV$^{-2}$ 
for the $\pi J/\psi$ and 
$\mathcal{N}_{D \bar{D}^{*}} |C_{Y \to \pi (\pi J/\psi)}|=1.04(41)\times 10^{-3}$ MeV$^{-2}$ 
for the $D \bar{D}^{*}$ invariant mass distributions, 
where $\mathcal{N}_{\alpha}$ are the normalization factors to the raw data.

Resulting invariant mass spectra are shown in Fig.~\ref{fig8} 
where the shaded bands denote the statistical errors:
we observe that the invariant mass spectra calculated 
with the coupled-channel potential $V^{\alpha \beta}$ 
well reproduce the peak structures just above the $D \bar{D}^{*}$ threshold.
In both Figs.~\ref{fig8} (a) and \ref{fig8} (b),
the peak positions around $3.9$ GeV are denoted by vertical arrows.
We also find that the reflection peak in Fig.~\ref{fig8} (a) 
due to the symmetrization of two identical pions.
If we turn off the off-diagonal components of $V^{\alpha \beta}$ with the same constants 
$C_{Y \rightarrow \pi+\alpha}$, we obtain the results shown by the dashed lines,
where the lines are normalized to the results obtained from the full calculations at $4$ GeV.
The peak structures around $3.9$ GeV can not be reproduced without the off-diagonal components.

\section{Summary}
\label{summary}

We have studied the $\pi J/\psi$-$\rho \eta_c$-$D \bar{D}^{*}$ coupled-channel interactions
using (2+1)-flavor full QCD gauge configurations,
in order to unravel the structure of the tetraquark candidate $Z_c(3900)$.
Thanks to the HAL QCD method, we obtain the full coupled-channel potential $V^{\alpha \beta}$, 
whose diagonal components are all small.
This indicates that 
the $Z_c(3900)$ cannot be a simple hadro-charmonium or $D \bar{D}^{*}$ molecular state.

Also, we have found the transition potential between  $\pi J/\psi$ and $D \bar{D}^{*}$ is strong,
which indicates that the $Z_c(3900)$ can be explained as a threshold cusp.
To confirm this, we calculate the invariant mass spectra and pole positions associated with the coupled-channel two-body S-matrix on the basis of  $V^{\alpha \beta}$.
The results indeed support that the peak in the $\pi J/\psi$ invariant mass spectrum is not associated with a conventional resonance but is a threshold cusp induced by the strong $\pi J/\psi$-$D \bar{D}^{*}$ coupling.
To further strengthen our conclusion, we have made a semiphenomenological analysis of the three-body decay of the $Y(4260)$, and find that the experimentally observed peak structures just above the $D \bar{D}^{*}$ threshold are well reproduced in the $Y(4260) \to \pi \pi J/\psi$ and the $Y(4260) \to \pi D \bar{D}^{*}$ decays.

To make a definite conclusion on the structure of the $Z_{c}(3900)$ in the real world,
we plan to carry out full QCD simulations near the physical point.
It is also an interesting future problem to study the structure of 
pentaquark candidates $P_c^+(4380)$ and $P_c^+(4450)$ on the basis of the coupled-channel HAL QCD method.

\ack
The author thanks all the member of the HAL QCD Collaboration for discussion.
The author is also grateful 
to ILDG/JLDG~\cite{JLDG} for providing us with full QCD gauge configurations used in this study, 
and to Doctor C.Z. Yuan for providing us with BESIII experimental data.
Numerical calculations were carried out on NEC-SX9 at RCNP in Osaka University and SR16000 at YITP in Kyoto University.
This study is supported in part by JSPS KAKENHI Grants Number JP17K14287,
and by MEXT as ``Priority Issue on Post-K computer'' (Elucidation of the Fundamental Laws and Evolution of the Universe)
and SPIRE (Strategic Program for Innovative REsearch).


\appendix

\section{Complex poles of the coupled-channel S-matrix}
\label{app:pole}
\begin{figure*}[htb]
\begin{center}
\includegraphics[width=0.39\textwidth,clip]{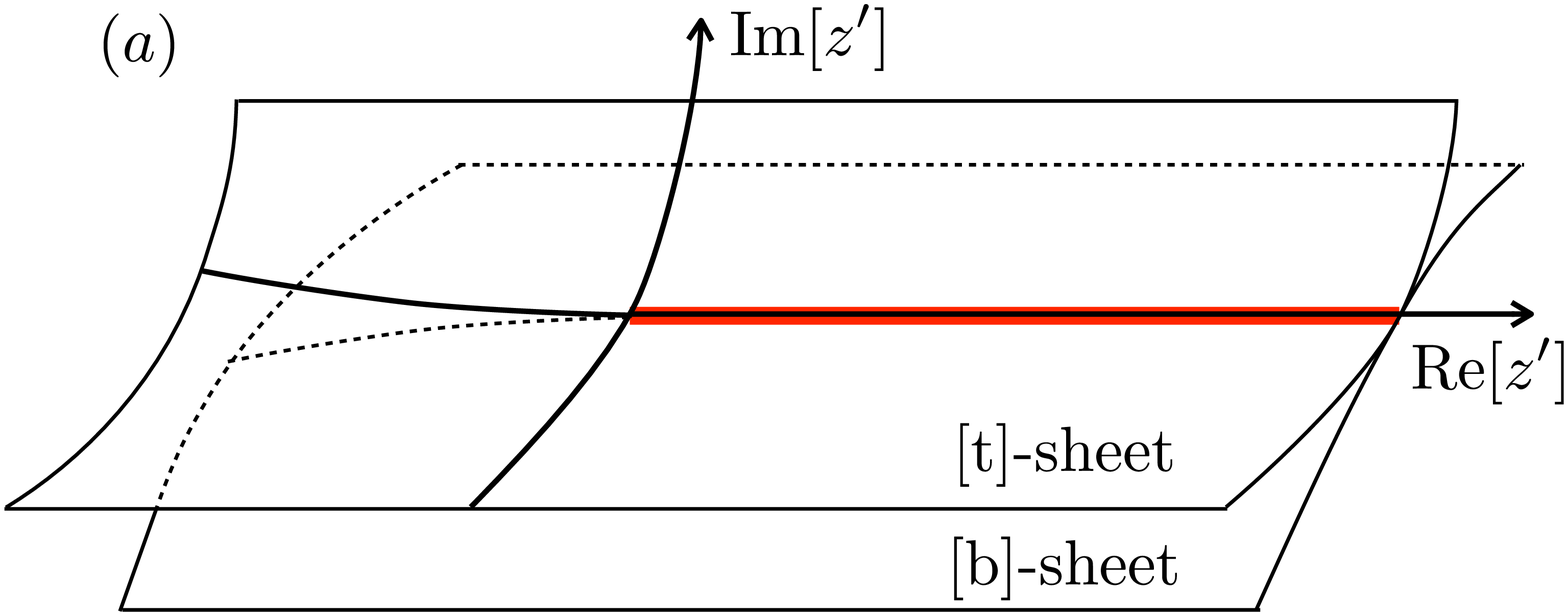}
\includegraphics[width=0.39\textwidth,clip]{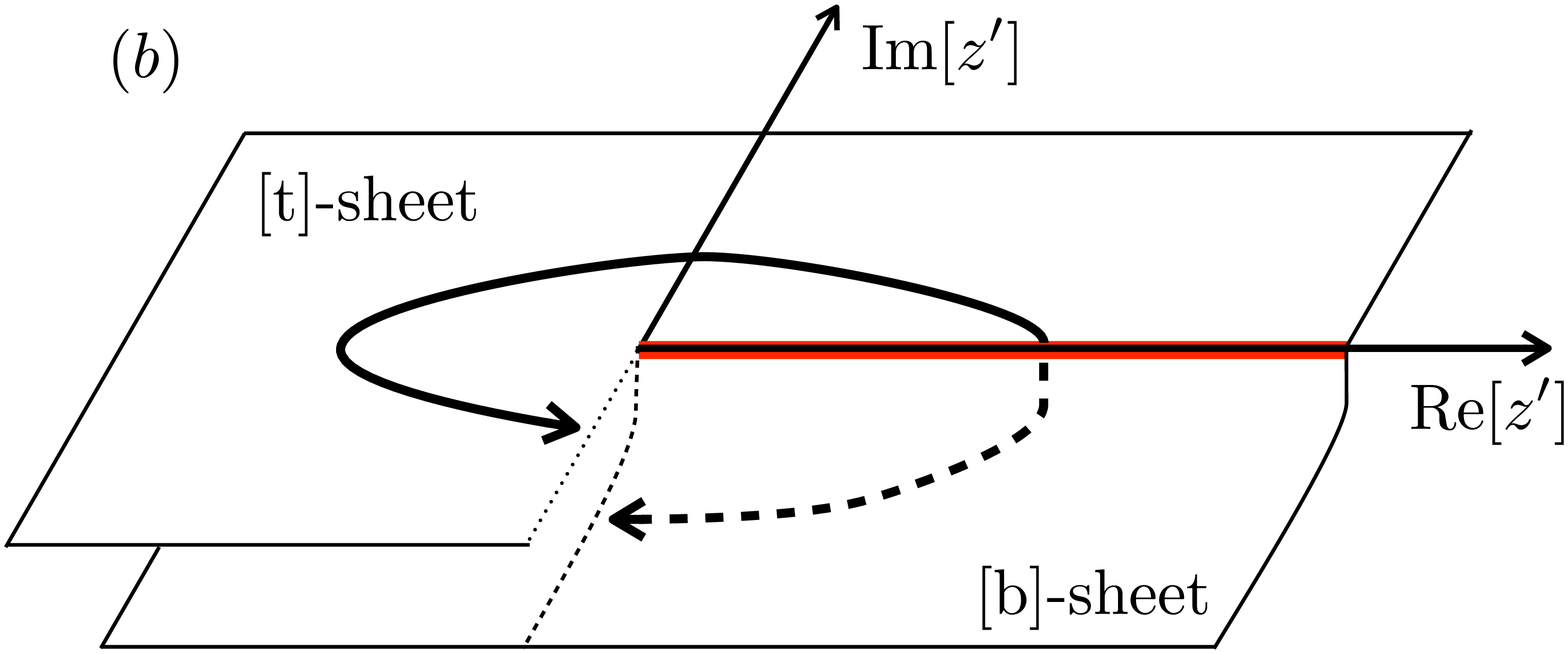}
\includegraphics[width=0.19\textwidth,clip]{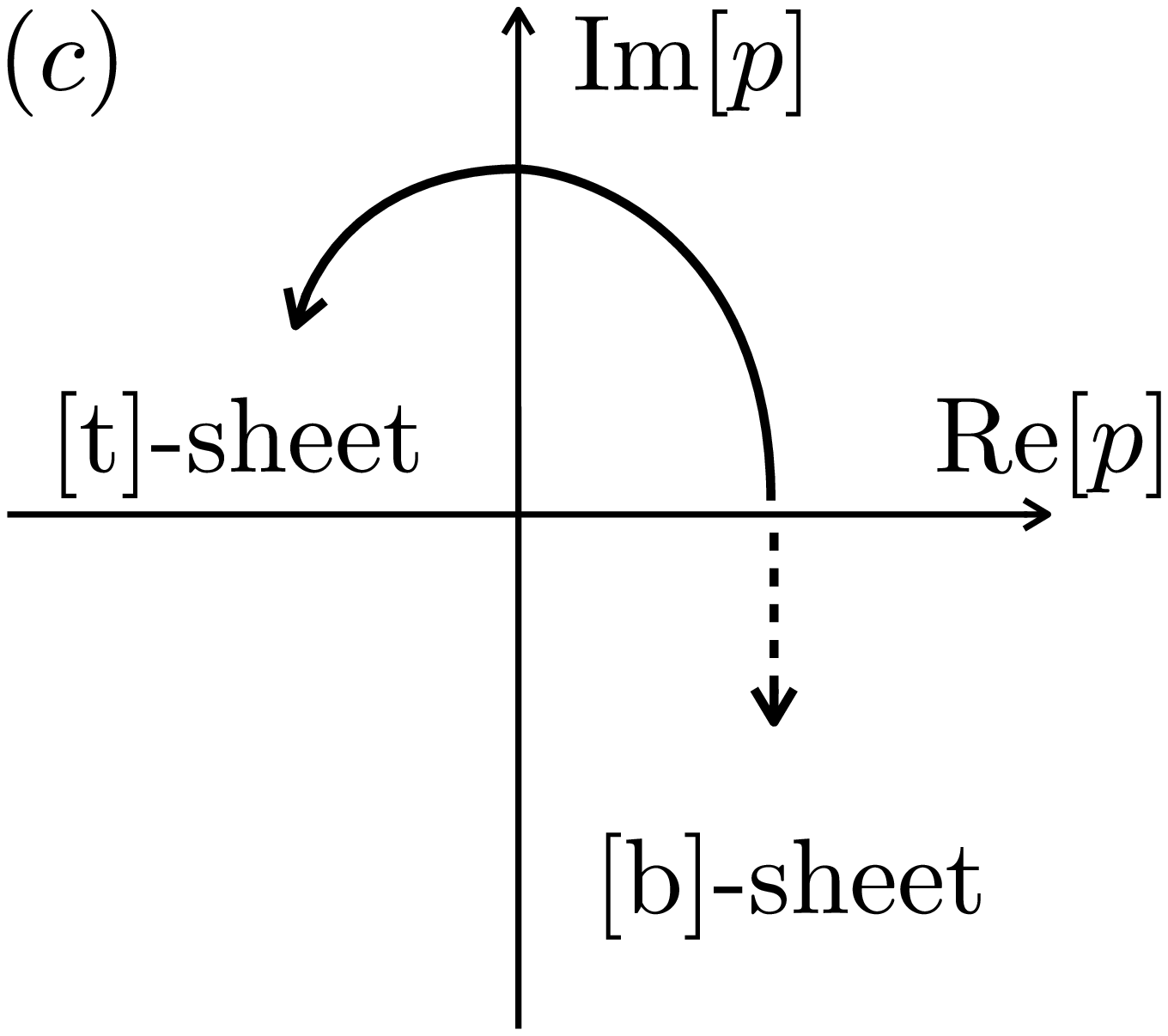}
\end{center}
\caption{
Structure of Riemann sheets for single-channel case with the notation of Ref.~\cite{Pearce:1988rk}. 
The complex energy $z$ is related to the complex momentum $p$ by $z \equiv m_1+m_2+p^2/(2\mu)$
with $\mu$ being the reduced mass.
In Fig.~\ref{fig:A1} (a) with $z' \equiv z- (m_1+m_2)$, 
the upper-half of the [$t$]-sheet is continuously connected to the lower-half [$b$]-sheet across the branch cut denoted by the red line.
In Fig.~\ref{fig:A1} (b), which is equivalent to Fig.~\ref{fig:A1} (a), 
is used for illustrating the Riemann sheets in the coupled-channel case 
(Figs.~\ref{fig5} and \ref{fig:A3}). 
The [$t$]-sheet and the [$b$]-sheet correspond to 
the upper-half and the lower-half of the complex $p$-plane as shown in Fig.~\ref{fig:A1} (c). 
}
\label{fig:A1}
\end{figure*}

\begin{figure*}[b]
\begin{center}
\includegraphics[width=0.75\textwidth,clip]{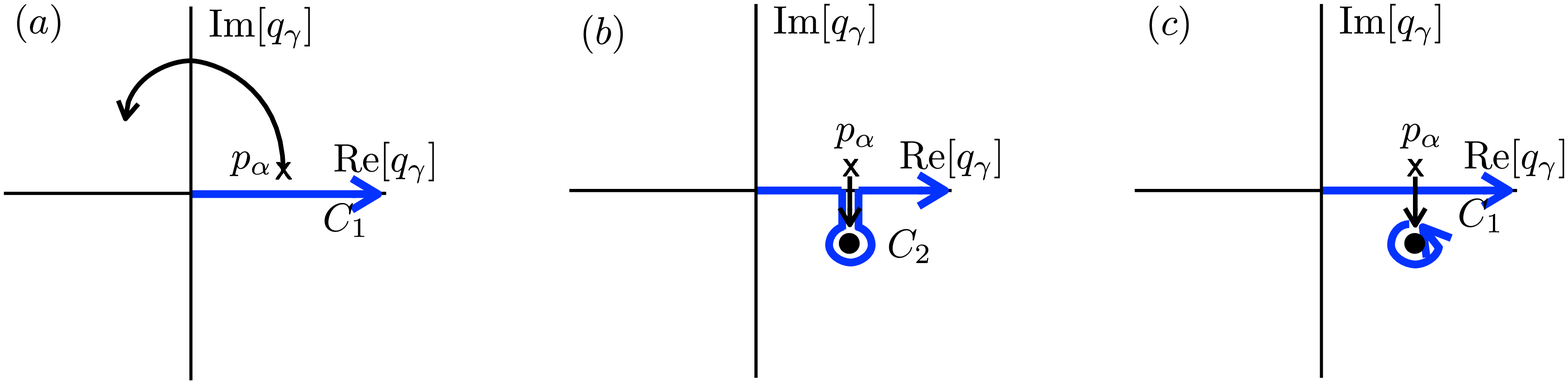}
\end{center}
\caption{
The integration contours in the complex momentum plane of $q_{\gamma}$.
The contour $C_{1}$ lies on the real axis in Figs.~\ref{fig:A2} (a) and \ref{fig:A2} (c), 
while the contour $C_{2}$ is deformed into the forth quadrant in the $q_{\gamma}$ plane 
due to the analytic continuation as shown in Fig.~\ref{fig:A2} (b).
We note that Figs.~\ref{fig:A2} (b) and \ref{fig:A2} (c) correspond to 
Fig. 1 (d) in Ref.~\cite{Pearce:1988rk}.
}
\label{fig:A2}
\end{figure*}

In this Appendix, we briefly describe the method of the analytic continuation 
of two-body amplitudes onto the complex energy plane following Ref.~\cite{Pearce:1988rk}.
After the $s$-wave projection, 
the coupled-channel Lippmann-Schwinger (LS) equation in Eq.~(\ref{eq:2-body.LS_eq}) 
for the s-wave two-body T-matrix $t_{(\ell = 0)}^{\alpha \beta}$ 
($\alpha, \beta = \pi J/\psi, \rho \eta_c, D \bar{D}^{*}$) reads
\begin{eqnarray}
\fl
t^{\alpha \beta}_{(\ell=0)}(p_{\alpha}, p_{\beta}; E)
=
V^{\alpha \beta}_{(\ell=0)}(p_{\alpha}, p_{\beta})
+ 
\sum_{\gamma} \int d q_{\gamma} q_{\gamma}^{2}
\frac{ V^{\alpha \gamma}_{(\ell=0)}(p_{\alpha}, q_{\gamma}) 
t^{\gamma \beta}_{(\ell=0)}(q_{\gamma}, p_{\beta}; E) }
{ E - E_{\gamma}(q_{\gamma}) + i \epsilon } 
~,
\label{eq:A.LS_s-wave}
\end{eqnarray}
where $p_{\alpha,\beta} = \left| \vec{p}_{\alpha,\beta} \right|$ and 
$q_{\gamma} = \left| \vec{q}_{\gamma} \right|$ are 
the real on-shell (off-shell) momentum of the two-meson state in channel $\alpha$ ($\gamma$);
The scattering energy in the center-of-mass frame is
$E = m_1^{\alpha} + m_2^{\alpha} + \vec{p}^2_{\alpha}/2\mu^{\alpha}$, while 
$E_{\gamma}(\vec{q}_{\gamma}) = m_1^{\gamma} + m_2^{\gamma} + \vec{q}^2_{\gamma}/2\mu^{\gamma}$
represents the energy of the intermediate states in channel $\gamma$.
Following Eq.~(\ref{eq:2-body.S-mat}), the relation to the on-shell S-matrix is given as
\begin{eqnarray}
S^{\alpha \beta}_{(\ell=0)}(p_{\alpha}, p_{\beta}; E)
& = &
\delta_{\alpha \beta}
- 2 \pi i
\sqrt{ \mu^{\alpha} p_{\alpha}  \mu^{\beta} p_{\beta} }
\ t^{\alpha \beta}_{(\ell=0)}(p_{\alpha}, p_{\beta}; E) ~.
\nonumber
\end{eqnarray}

To study the pole structure of the $Z_{c}(3900)$ in the complex energy plane,
we need to carry out  analytic continuation of Eq.~(\ref{eq:A.LS_s-wave}) 
in terms of  the complex energy $z \equiv m_1^{\alpha} + m_2^{\alpha} + p_{\alpha}^2/2\mu^{\alpha}$
with $p_{\alpha}$ being the complex momentum.
The top sheet and the bottom sheet 
(or the [$t$]-sheet and the [$b$]-sheet according to the notation of Ref.~\cite{Pearce:1988rk})
are joined along the branch cut on the real axis 
starting from $z' \equiv z - (m_1^{\alpha} + m_2^{\alpha}) =0$ 
as shown in Fig.~\ref{fig:A1} (a). 
An alternative but equivalent way of illustrating the same structure is 
given by Fig.~\ref{fig:A1} (b) which is found to be more useful in multi-channel cases.
Fig.~\ref{fig:A1} (c) shows the complex-momentum plane 
where the [$t$]-sheet and the [$b$]-sheet correspond to 
$0 \le \arg \left( p_{\alpha} \right) < \pi$ and 
$\pi \le \arg \left( p_{\alpha} \right) < 2\pi$, respectively.

We define the analytic continuation of the integral in Eq.~(\ref{eq:A.LS_s-wave}) as $I^{\gamma}(z)$.
For $z$ located in the [$t$]-sheet, the integral can be carried out 
by choosing  the contour $C_1$ for $q_\gamma$-integration (see Fig.~\ref{fig:A2} (a)):
\begin{eqnarray}
I_{\rm [t]}^{\gamma}(z)
& = &
\int_{C_{1}} d q_{\gamma} q_{\gamma}^{2}
\frac{ V^{\alpha \gamma}_{(\ell=0)}(p_{\alpha}, q_{\gamma}) 
t^{\gamma \beta}_{(\ell=0)}(q_\gamma, p_{\beta}; z) }
{ z - E_{\gamma}(q_{\gamma}) } ~.
\nonumber
\end{eqnarray}
On the other hand, for $z$ located in the [$b$]-sheet, the integration contour 
should be chosen to be $C_2$ for analytic continuation (see Fig.~\ref{fig:A2} (b)).
This is equivalent to picking up the anti-clockwise residue 
at the pole $+$ the integration along the contour $C_1$ 
as shown in Fig.~\ref{fig:A2} (c):
\begin{eqnarray}
\fl
I_{\rm [b]}^{\gamma}(z)
& = &
\int_{C_{2}} d q_{\gamma} q_{\gamma}^{2}
\frac{ V^{\alpha \gamma}_{(\ell=0)}(p_{\alpha}, q_{\gamma}) 
t^{\gamma \beta}_{(\ell=0)}(q_{\gamma}, p_{\beta}; z) }
{ z - E_{\gamma}(q_{\gamma}) } 
\nonumber \\
\fl
& = &
\int_{C_{1}} d q_{\gamma} q_{\gamma}^{2}
\frac{ V^{\alpha \gamma}_{(\ell=0)}(p_{\alpha}, q_{\gamma}) 
t^{\gamma \beta}_{(\ell=0)}(q_{\gamma}, p_{\beta}; z) }
{ z - E_{\gamma}(q_{\gamma}) }
- 2\pi i \mu^{\gamma} p_{\gamma} 
V^{\alpha \gamma}_{(\ell=0)}(p_{\alpha}, p_{\gamma})
t^{\gamma \beta}_{(\ell=0)}(p_{\gamma}, p_{\beta}; z) ~.
\nonumber
\end{eqnarray}
Thus,
we obtain a coupled-channel LS equation 
defined on the complex energy plane~\cite{Pearce:1988rk}:
\begin{eqnarray}
t^{\alpha \beta}_{(\ell=0)}(p_{\alpha}, p_{\beta}; z)
& = &
V^{\alpha \beta}_{(\ell=0)}(p_{\alpha}, p_{\beta})
+ \sum_{\gamma} I_{\rm [t, b]}^{\gamma}(z) ~.
\label{eq:CCC}
\end{eqnarray}

\begin{figure*}[htb]
\begin{center}
\includegraphics[width=0.60\textwidth,clip]{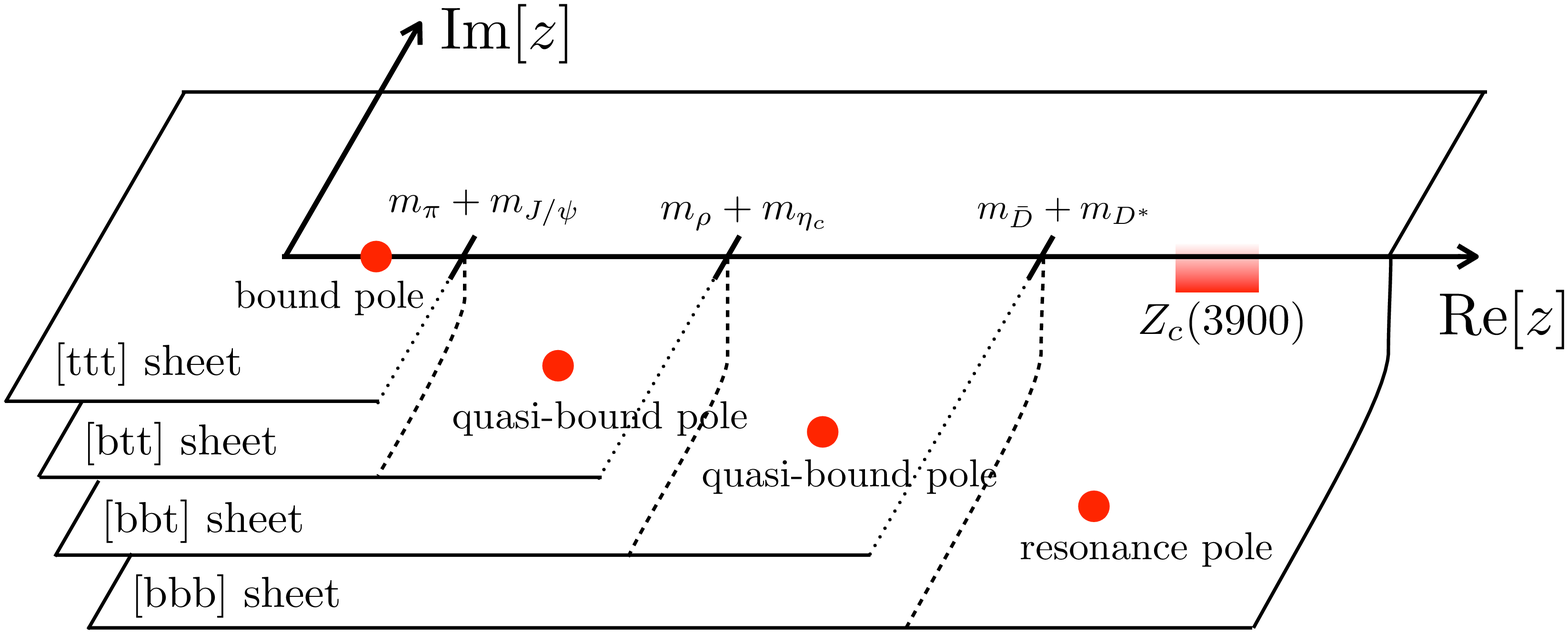}
\end{center}
\caption{
The complex energy plane 
in the $\pi J/\psi$-$\rho \eta_c$-$D \bar{D}^{*}$ coupled-channel system.
The energy  relevant to $Z_c(3900)$ is indicated by shaded area, and
 examples of a possible resonance pole,  quasi-bound poles and a bound pole are 
 illustrated by the red filled circles.}
\label{fig:A3}
\end{figure*}

With Eq.~(\ref{eq:CCC}) for the $\pi J/\psi$-$\rho \eta_c$-$D \bar{D}^{*}$ system,
pole positions of the coupled-channel S-matrix are examined on 8 Riemann sheets originating from the existence of three thresholds.
These sheets are characterized by the notation [$xyz$]  with $x$, $y$ and $z$ taking either $t$ or $b$, according to Ref.~\cite{Pearce:1988rk}.
Among 8 sheets, the poles near the real axis on the [$bbb$], [$bbt$], [$btt$] and [$ttt$] sheets are most relevant for the scattering observables,
since these sheets are directly connected to the physical region.
On the other hand, the poles on the [$tbt$], [$ttb$], [$tbb$] and [$btb$] sheets hardly affect the  scattering observables, 
since they are not directly connected to the physical region.
In Fig.~\ref{fig:A3}, we show some examples of ``possible'' poles on the [$bbb$], [$bbt$], [$btt$] and [$ttt$] sheets: 
 a $D \bar{D}^{*}$ resonance  pole near the real axis on the [$bbb$] sheet,   
 a $D \bar{D}^{*}$ quasi-bound  pole  near the real axis on the [$bbt$] sheet, 
 a $\rho \eta_{c}$  quasi-bound pole  near the real axis on the [$btt$] sheet,
and a $\pi J/\psi$ bound pole on the [$ttt$] sheet below all the thresholds.
The energy region relevant to $Z_c(3900)$ is shown by the shaded area in Fig.~\ref{fig:A3}.

\begin{figure*}[t]
\begin{center}
\includegraphics[width=0.40\textwidth,clip]{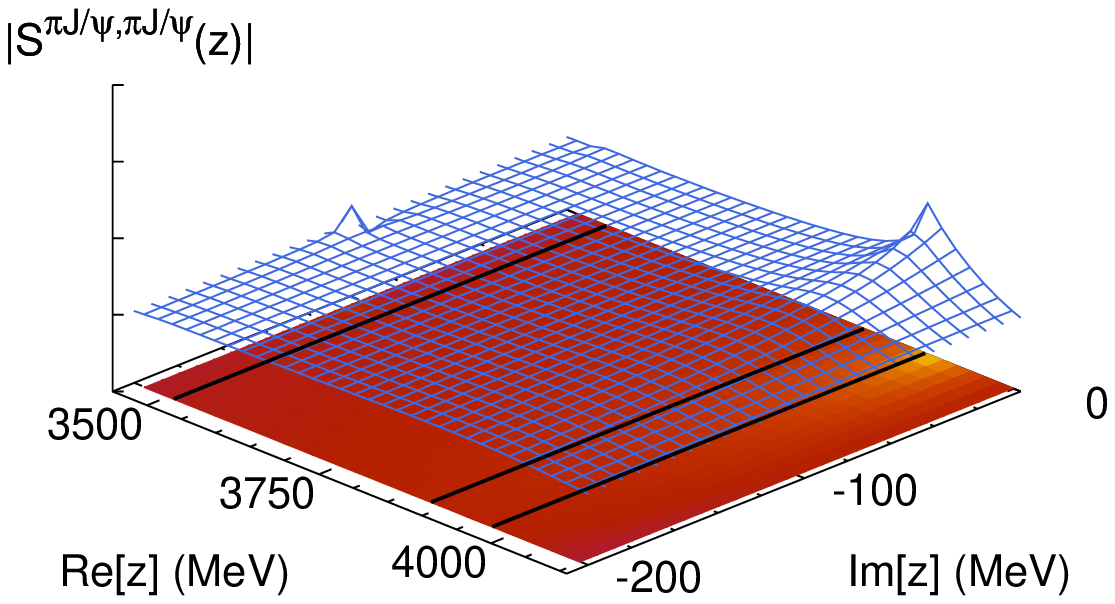}
\put(-180,120){(a)}
\hspace{1cm}
\includegraphics[width=0.40\textwidth,clip]{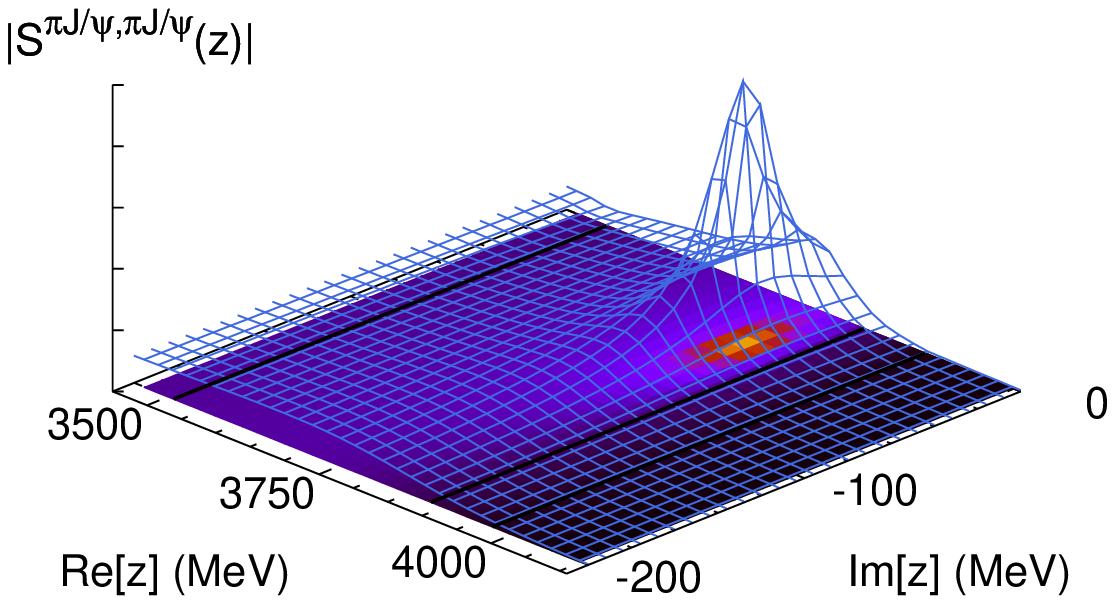}
\put(-180,120){(b)}\\
\includegraphics[width=0.40\textwidth,clip]{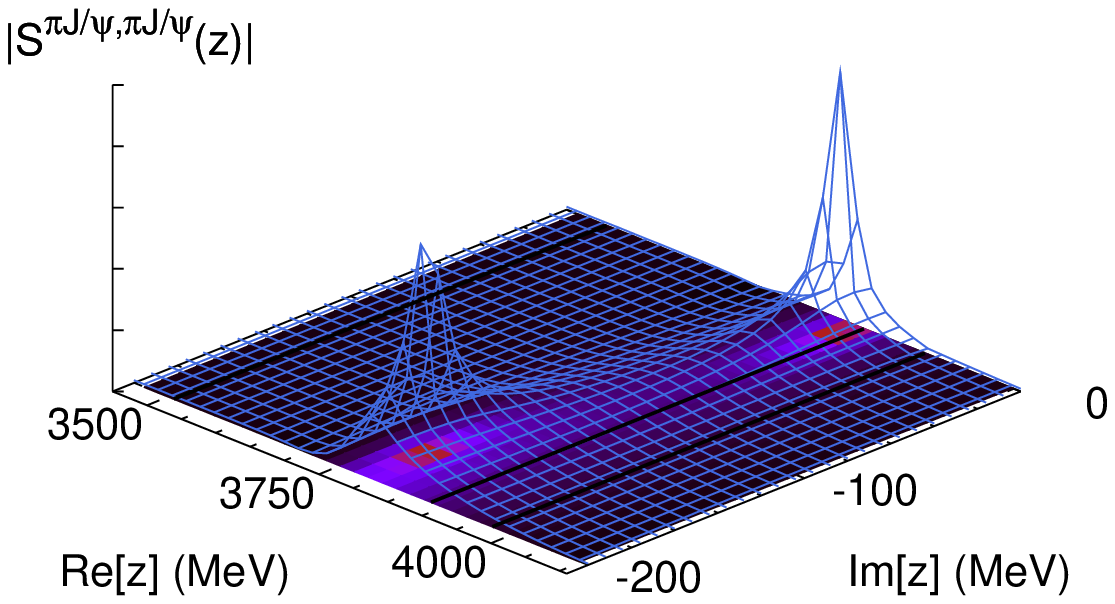}
\put(-180,120){(c)}
\hspace{1cm}
\includegraphics[width=0.40\textwidth,clip]{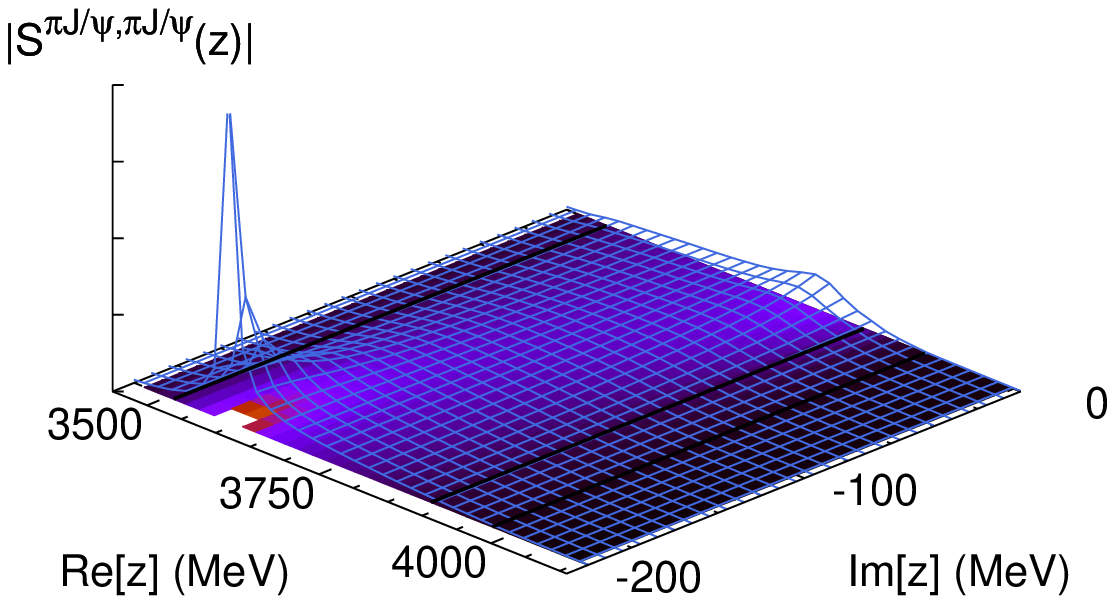}
\put(-180,120){(d)}
\end{center}
\caption{
The absolute magnitude of the S-matrix in the $\pi J/\psi$-$\pi J/\psi$ channel
on the (a) [$tbt$], (b) [$ttb$], (c) [$tbb$] and (d) [$btb$] sheets
in the notation of Ref.~\cite{Pearce:1988rk}.
The quark mass corresponds to case I of Table~\ref{tab1} in the main text.
}
\label{fig:A4}
\end{figure*}

In the main text, we have numerically searched through the poles 
on all four relevant sheets in the $\pi J/\psi$-$\pi J/\psi$ S-matrix,
and we have not found a pole corresponding to a bound or a quasi-bound state on the [$ttt$], [$btt$] and [$bbt$] sheets.
Meanwhile, on the [$bbb$] sheet, 
we have found a pole far below the $D \bar{D}^{*}$ threshold with a large imaginary part,
so that the pole does not affect scattering observables.

Just for completeness, we show, in Fig.~\ref{fig:A4},  
the absolute value of the S-matrix on the (a) [$tbt$], (b) [$ttb$], (c) [$tbb$] and (d) [$btb$] sheets for case I of Table~\ref{tab1}.
We find poles on the [$ttb$], [$tbb$] and [$btb$] sheets, 
although they do not affect the observables as we mentioned before. 
The numerical results for the pole positions not only for case I but also for case II and III 
are summarized in Table~\ref{tab:A1}.
In Table~\ref{tab:A1}, the mean value is calculated 
with the coupled-channel potential $V^{\alpha \beta}$ at $t=13$.
The first parenthesis indicates the statistical error from lattice QCD data,
and the second parenthesis indicates the systematic error 
evaluated by the difference between the pole position at $t=13$ and that at $t=15$.

\begin{table*}[!h]
   \begin{tabular}{c|c|c|c}
      \hline
      \hline
    & [$ttb$] & [$tbb$] & [$btb$] \\
      \hline 
I   &  {\small $-146(112)(108) - i 38(148)(32)$} & {\small $-177(116)(61) - i 175(30)(22)$} & {\small $-369(129)(102) - i 207(61)(20)$}  \\
&  {\small $-93(55)(21) - i 9(25)(7)$}  & \\
      \hline 
 II  & {\small $-102(84)(45) - i 14(11)(7)$} & {\small $-141(92)(64) - i 151(149)(132)$} & {\small $-322(141)(111) - i 114(96)(75)$} \\
&  {\small $-59(67)(11) - i 3(12)(1)$} & \\
      \hline 
 III & {\small $-100(48)(29) -i 7(37)(17)$} & {\small $-127(52)(43) -i 199(44)(28)$} & {\small $-356(108)(28) -i 277(138)(95)$} \\
&  {\small $-53(30)(5) -i 2(11)(3)$} & \\
      \hline 
      \hline 
   \end{tabular}
   \caption{
   The pole positions $z_{\rm pole} - m_{D} - m_{\bar{D}^{*}}$ in MeV unit
   in different sheets with three cases for the pion mass.
   Case I, II and III correspond to 
   those of Table~\ref{tab1} in the main text.
   The mean value is calculated with the potential at $t=13$.
   The first parenthesis indicates the statistical error from lattice data,
   and the second parenthesis indicates the systematic error 
   from the difference between the pole position at $t=13$ and that at $t=15$.
     }
   \label{tab:A1}
\end{table*}

\end{document}